\documentclass[useAMS,usenatbib]{mn2e}
\title[The evolution of the galaxy B-band rest-frame morphology to $z\sim 2$]{The evolution of the galaxy B-band rest-frame morphology to $z\sim 2$: new clues from the K20/GOODS sample}
\author[P.Cassata et al.]
{P.~Cassata$^{1}$\thanks{E-mail: cassata@pd.astro.it},
A.~Cimatti$^{2}$, A.~Franceschini$^{1}$, 
E.~Daddi$^{5}$, E.~Pignatelli$^{3}$,\newauthor 
G.~Fasano$^{3}$, G.~Rodighiero$^{1}$, L.~Pozzetti$^{4}$, 
M.~Mignoli$^{4}$ and A.~Renzini$^{5}$
\\
$^{1}$Dipartimento di Astronomia, Vicolo Osservatorio 2, I-35122, Padova, Italy\\
$^{2}$INAF, Osservatorio Astronomico di Arcetri, Largo E. Fermi 5, I-50125 Firenze, Italy\\
$^{3}$INAF, Osservatorio Astronomico di Padova, Vicolo Osservatorio 2, I-35122, Padova, Italy\\
$^{4}$INAF, Osservatorio Astronomico di Bologna, Via Ranzani 1, I-40124, Bologna, Italy\\
$^{5}$European Southern Observatory, Karl-Schwarzschild-Str. 2, D-85748, Garching, Germany\\
}
\usepackage{natbib}
\usepackage{graphicx}
\usepackage{graphics}
\usepackage{amssymb}
\usepackage{amsmath}
\usepackage{fancyheadings}
\usepackage{longtable}
\usepackage{float}

\begin{document}

\date{}

\pagerange{\pageref{firstpage}--\pageref{lastpage}} \pubyear{2004}

\hyphenation{ana-lysis} 

\maketitle

\label{firstpage}

\begin{abstract}
We present a detailed analysis of the evolution of the rest-frame 
B-band morphology of K-selected galaxies with $0<z<2.5$. 
This work is based on the K20 spectroscopic sample ($Ks<20$) located 
within the Chandra Deep Field South area, coupled with the public deep 
GOODS HST+ACS multi-band optical imaging available in that field.
Thanks to the spectroscopic completeness 
of this catalog reaching 94\%, we can compare the morphological and 
spectroscopic properties of galaxies with unprecedented detail.
Our morphological analysis includes visual inspection and automatic procedures 
using both parametric (e.g. the S\'ersic indices treated by the 
GALFIT and GASPHOT packages) and non-parametric (the Concentration, Asymmetry 
and clumpineSs, CAS) methods.
By exploiting the 4-band deep ACS imaging we account in detail for the 
morphological K-correction as a function of the redshift and show that, 
while the parametric methods do not efficiently separate early- and 
late-type galaxies, non-parametric ones prove more efficient and reliable. 
Our analysis classifies the K20 galaxies as: 60/300 (20\%, class 1) normal 
ellipticals/S0; 
14/300 (4\%, class 2) perturbed or peculiar ellipticals; 80/300 
(27\%, class 3) normal spirals; 48/300 (16\%, class 4) perturbed or actively 
star-forming spirals;  98/300 (33\%, class 5) irregulars. 
The morphological and spectroscopic classifications are compatible with each 
other for more than 90\% of the sample galaxies, while 7 class-1 E/S0's 
show emission lines and 11 spirals and irregulars (class 3+4+5) have purely 
absorption-line spectra. 
The evolution of the merging fraction is constrained up to $z\sim2$, by carefully
accounting effects of morphological $k$-correction: both asymmetry criterion
and pair statistic show an increasing merging fraction as a function of the
redshift.
We finally analyse the redshift-dependence of the 
effective radii for early- and late-type galaxies and find some mild evidence 
for a decrease with $z$ of the early-type galaxy sizes, while disks and 
irregulars remain constant.
Altogether, this analysis of the K20 sample indicates the large predominance 
of spirals and irregulars at $0.5<z<1.5$ in K-band selected samples at even 
moderate depths.
\end{abstract}

\begin{keywords}
galaxies: evolution -- galaxies: interactions -- galaxies: structure
\end{keywords}

\section{Introduction}

Galaxies in the local universe can be organized in a sequence of 
morphologies
(e.g. the Hubble sequence) which must be the result of the specific 
processes
having originated them. Yet, the relative roles over cosmic time
of processes such as merging
of dark matter halos, dissipation, starburst, feedback, AGN activity, etc.
remain largely conjectural, in particular concerning the establishment 
of the galaxy morphological differentiation. Therefore, morphological 
analyses of faint high-z galaxies and studies of the 
evolution of galaxy sizes with cosmic time give an important insight on how 
the matter aggregated into the structures that we see today. 
The combination of a morphological investigation for flux limited samples 
of faint galaxies with complete redshift information provides decisive 
constraints on the formation epoch and the pattern for the galaxy build-up.

While in the local universe approximately 65\% of all galaxies are spirals,
32\% are ellipticals and 4\% irregulars/peculiars (Marzke et al. 1998),
there are indications that in the distant universe the irregular/peculiar 
fraction becomes predominant (i.e. Glazebrook et al. 1995, Abraham 1996).
Van den Bergh et al. (2000, 2001) concluded that most
of the galaxies with $z\gtrsim0.5$ can be hardly classified within the Hubble 
scheme: irregular/peculiar/merging objects become more common beyond $z\sim0.3$,
spiral structures at $z\gtrsim0.6$ are more chaotic than locally and 
spirals and ellipticals become extremely rare at $z\gtrsim1.5$.
Conselice et al. (2004) reached similar conclusions, adding that at
$z\gtrsim2$ over 80\% of the stellar mass is stored in peculiar galaxies. 

Exploiting the same data set used in this work Cimatti et al. (2003) investigated 
the morphology of the K20/CDFS EROs, finding that only the $\sim30\%$ of these 
objects were early-type galaxies.

As a continuation of the K20 survey, Daddi et al. (2004) identified 9
actively star-forming/merging  galaxies in the K20/GOODS field with
$1.7<z<2.3$ and stellar mass $M_*>10^{11}M_\odot$, while Cimatti et al. 
(2004)
found over the same field 4 passive, early-type galaxies with $1.6<z<1.9$
with similar stellar masses. Thus, while at $z=0$ most so massive galaxies
are elliptical/S0 galaxies, by $z\sim 2$ starburst and passive galaxies
appear to be in nearly equal numbers.

The Cold Dark Matter (CDM) models of galaxy formation predict that
dark matter halos formed by the merging of smaller units in the past.
Thus, the evolution of the merging fraction with redshift gives also important
cosmogonic information (see for a review Abraham 1998).
%In the local universe galaxies are usually classified according the Hubble
%system, that is known to partly lose significance when dealing with samples 
%obtained by medium and deep surveys.
Several studies report an increasing merging fraction
with redshift, both at moderate ($z<0.3$, Patton 
et al. 1997), and at higher redshifts (Le F\`evre et al. 2000, by pair counts;
Conselice et al. 2003, by the Asymmetry measurements).

However, in order to study the evolution with redshift of galaxy morphologies and 
sizes, the dependence of morphological properties on the wave-band and 
the effects of the changing rest-frame wavelength as a function of redshift 
(the so-called morphological $K$-correction) have to be taken into account. 
Galaxies observed in their blue rest-frame appear with later-type morphologies 
than observed in the red part of the spectrum (e.g. Windhorst et. al 2002; 
Papovich et al. 2003).

The problem of quantifying morphological properties of high redshift galaxies 
is still quite open.
In the literature, parametric and non-parametric methods are usually applied. 
The former attempt to model the light distributions with a combination of 
analytic laws, like de Vaucouleurs or exponential profiles, or the S\'ersic 
model (GALFIT, Peng et al. 2002 and GIM2D, Simard et al. 2002). 
Various parameters (i.e. bulge/disk B/D ratios or S\'ersic indices) are then 
derived, which are known to correlate with qualitative Hubble classifications. 
This approach suffers however a number of problems. It needs to assume that 
the galaxy light distribution is well reproduced by a symmetric 
profile, hence is not suitable to treat merging structures, spiral arms, dust 
lanes and so on.
Moreover, there are degeneracies in the solutions for 
multi-component fits (e.g. needed to retrieve the B/D ratio), because of 
the large number of free parameters. This becomes quickly unmanageable at 
decreasing galaxy brightness and number of galaxy pixels.

An often used representation of galaxy morphological types is the parameter 
$n$ associated to the S\'ersic profile $\mu \propto exp(-1/n)$, with some 
clear advantages over multi-component fits (Pignatelli et al. 2004). 
Unfortunately, there are even local ellipticals with a roughly exponential
profile, while only the most luminous and massive objects have $n\sim4$ 
(i.e. Caon et al. 1993). This obviously implies an intrinsic level of degeneracy.

\begin{figure*}
\begin{center}
\includegraphics[width=\textwidth]{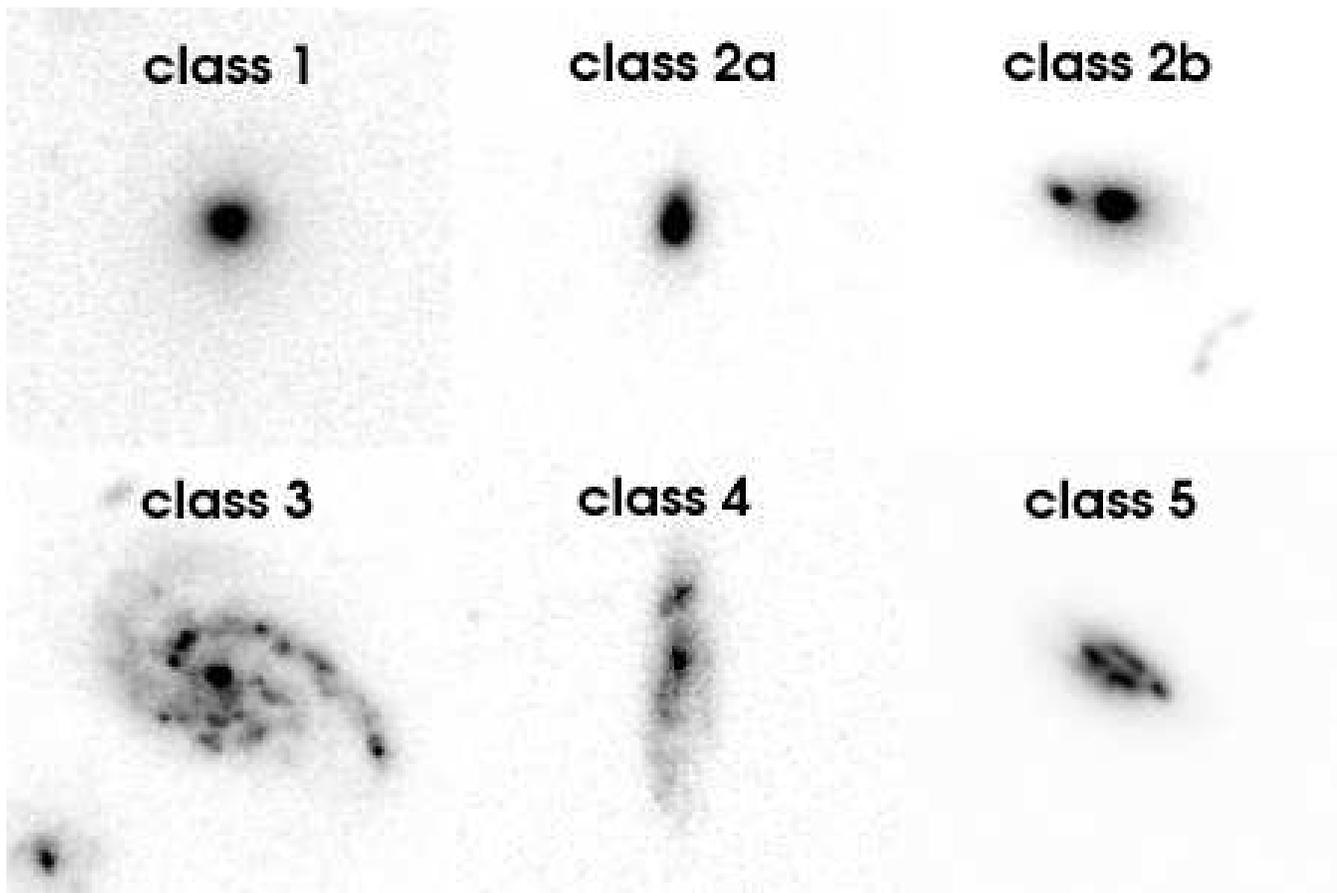}
\end{center}
\caption{
Some examples of galaxies representing morphological classes. Each
box measures $4.5''\times4.5''$. From
top-left to bottom-right we report: an isolated elliptical (class 1);
a perturbed elliptical in which a little distortion of isophotes can 
be appreciated (class 2a); a couple in which main object is an elliptical
(class 2b); a normal spiral (class 3); a peculiar spiral, in which signs
of interaction with a little companion at bottom-right can be noted 
(class 4); an irregular galaxy (class 5).
}
\label{mor_ex} 
\end{figure*}

Alternatively, non-parametric approaches to quantitative morphology have 
been developed in the last few years by several authors. The most used ones
are the Concentration (C) parameter (Abraham et al. 1996), that roughly 
correlates with the S\'ersic index and with the B/D ratio, and the Asymmetry 
(A) parameter (Abraham et al. 1996, Conselice et al. 2000), able in principle 
to distinguish irregulars or mergers from more symmetric galaxies (E/S0/Sa). 

More recently another parameter has been introduced, the clumpiness (S),
measuring the degree of ``patchiness'' of a galaxy (i.e. the light
in the high spatial frequencies). The clumpiness is expected to correlate
with the star-formation rate (Conselice 2003). The advantage of the 
non-parametric approach is that no a-priory assumption is made about 
the distribution of light. 
Some problems however still remain, like the influence of the noise 
on the asymmetry and clumpiness measures,  or the choice of the spatial 
scale used to compute clumpiness, that must be adapted to the distance 
and the size of the object, or again the choice 
of the center of rotation for the computation of the asymmetry.

One aspect that must be taken in consideration when dealing with
a morphological analysis is the degree of human interaction needed to obtain the
fitting parameters, both for non-parametric and for parametric procedures,
in particular when treating large amounts of data, as those made available
by the new Advanced Camera for Surveys on HST (e.g. the GOODS, COSMOS, UDF 
survey projects to mention a few). The GALFIT tool (Peng et al. 2002), 
for example, has the advantage to allow masking out bad-pixel zones 
of the image (e.g. dust lanes or spiral arms), in order to improve 
the fit. However, this appears unfeasible object by object over very large 
samples.
GASPHOT, an automated tool recently developed by Pignatelli, Fasano and 
Cassata (2004), fitting the galaxy light-profile with a S\'ersic model, has 
been instead defined to retrieve in fully automatic mode fundamental 
photometric and morphological parameters (magnitudes, radii, 
axial ratios, S\'ersic indices) for all the objects in a given image. 

In this paper we exploit the very deep high resolution imaging recently 
obtained over the CDFS area with ACS/HST taken in the GOODS/HST
treasury program (Giavalisco et al. 2004) to make a careful 
morphological study of galaxies in the K20 sample within this area. 
In particular, we have exploited the multi-band coverage provided by HST
and the redshift information available for each object to minimize the 
effects of morphological $K$-correction: we have studied each object
in the ACS band closer to the B-band rest-frame.
We also take advantage of the K-band selection, that collects
the light from low-mass stars dominating the baryonic content of galaxies,
and thus providing a better mass completeness level. The K-band also allows 
to minimize the effects of the spectral $K$-correction, dust absorption, 
and evolution.

We are particularly motivated to such an extensive analysis by the excellent 
optical spectroscopic follow-up obtained for this sample with the 
ESO Very Large Telescope in the last several years. This paper is devoted 
to the study of galaxy morphologies and sizes as a function of redshift. 
A forthcoming paper will expand on statistical analyses of the galaxy 
distribution as a function of morphology and will attempt to interpret 
these data with modellistic representations.

The galaxy sample is presented in Sect. \ref{sample}. The HST/ACS data for 
the morphological analysis are presented in Sect. \ref{data}.
In Sect. \ref{analysis} we use parametric and non-parametric analyses, 
in particular with GASPHOT and GALFIT and the 
Concentration-Asymmetry-Clumpiness set, and compare their results with 
those of a visual inspection.
In Sect. \ref{frac} we discuss the evolution with the redshift of the
inferred fractions of each morphological class, including the merging fraction.
The high level of spectroscopic coverage ($\sim95\%$) is exploited for 
comparison with the morphological classification in Sect. \ref{mospe}, 
while our constraints on the redshift evolution of sizes for various 
morphological types are reported in Sect. \ref{sizes}. Our conclusion are 
drawn in Sect. \ref{conclusion}.

\section{The sample}\label{sample}

The original sample was selected in the K-band by the K20 team into two 
independent sky regions, one centered in the Chandra Deep Field 
South (CDFS) covering 32.2 $\rm{arcmin}^2$, the second centered around the 
QSO 0055-269, and covers 19.8 $\rm{arcmin}^2$.
Detailed informations can be found in \cite{Cimatti2002a}.

\begin{figure*}
\begin{center}
\includegraphics[height=0.8\textheight]{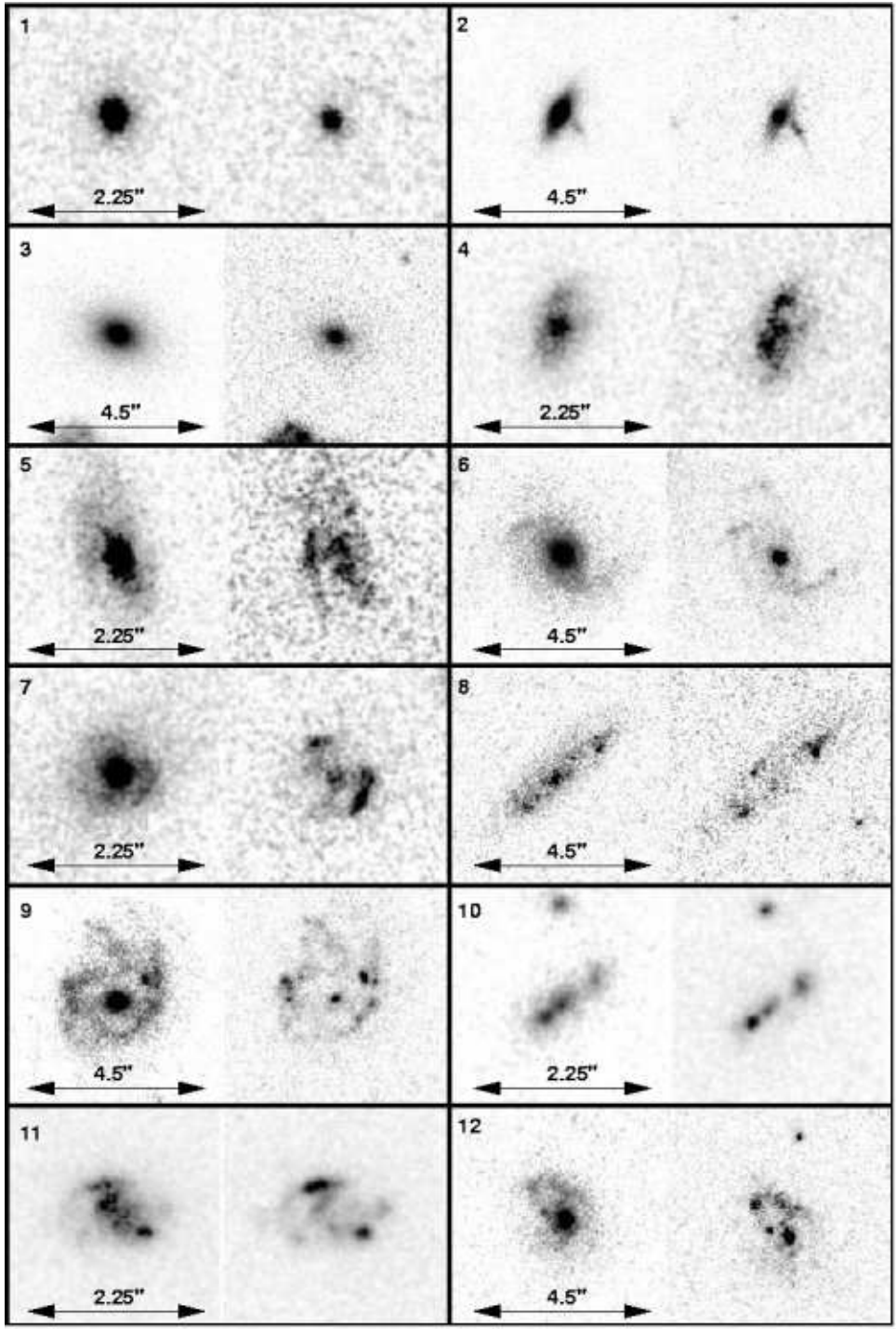}
\end{center}
\caption{
A sample of galaxies with $z\sim1$ are shown in their z-band, corresponding
to $\sim$4500 \AA\ rest-frame (left side of each panel), and in their V-band, 
corresponding to $\sim$3000 \AA\ (right side of each panel). These
examples are representative of morphological the $K$-correction problem
affecting galaxies observed in their U-band (that is galaxies having 
$z\gtrsim1.5$ observed in z-band) rather than in their B-band 
rest-frame. The panels labelled with 1, 2 and 3 contains elliptical/S0 
galaxies, panels with 4, 5 and 6 normal spirals, panels with 7, 8 and 9
perturbed spirals and panels with 10, 11 and 12 irregular/merger galaxies.
The size in arcsec of the images is reported in each panel.
}\label{kcorr} 
\end{figure*}

The complete sample consists of 546 objects (including a few stars) down to 
$K=20$. In this work, we restrict ourselves to the 346 objects lying in the 
CDFS area, for which high resolution imaging by ACS/HST has recently become 
available as a result of the GOODS HST Treasury program (Giavalisco et al. 2004).
After rejecting stars and quasars, our final sample includes 300 galaxies.

The available data include deep spectroscopy obtained with VLT+FORS1 and 
VLT+FORS2, and has been recently complemented with ESO/GOODS public data.
In the CDFS area the spectroscopic coverage is  $92\%$ with K20 data only,
and it reaches the $94\%$ including spectra by the ESO/GOODS public spectroscopic
survey (Vanzella et al. 2004).
For the remaining objects we used through out the paper the photometric redshifts 
derived using the ESO/$GOODS$ VLT+FORS1 $BVIRz$ and VLT+ISAAC $JHK_s$ 
public imaging (see Cimatti et al. 2002a). 

The sample includes objects belonging to two clusters at
$0.665<z<0.672$ and $0.732<z<0.740$ (see Cimatti et al. 2002b), including 
respectively 14 and 32 galaxies.

Galaxy spectra have been automatically classified according to their features 
into three main classes by Mignoli et al. (2004): 1. early type; 
2. early type + emission lines; 3. pure emission line.

\section{the HST/ACS data}\label{data}

Ground based photometry and spectroscopy was complemented with ACS imaging
in the $BViz$ bands taken in the $GOODS/HST$ Treasury Program 
(Giavalisco et al. 2003). We used the released version 1.0 of the images.
The GOODS ACS/HST Treasury Program has 
surveyed two separate fields (the Chandra Deep Field South and the
Hubble Deep Field North) with four broad band filters: F435W (B), F606W(V), 
F775W(i) and F850LP(z). 
Observations in the V, i and z filters have been split into 
5 epochs, separated by about 45 days, in order to detect transient objects.
Observations in the B band are taken during epoch 1 for both fields. 
Images taken at consecutive epochs have position angles increasing of 45 
degrees.
Total exposure times are 2.5, 2.5, 5 orbits in the V, i and z bands respectively. 
The exposure time in the B-band is 3 orbits.
In August 2003 the GOODS team released the version 1.0 of the reduced, stacked
and mosaiced images for all the data acquired during the five epochs of
observation. To improve the PSF sampling, the original images, which have
a scale of 0.05 arcsec/pixel, have been drizzled onto images with a scale
of 0.03 arcsec/pixel.

We have exploited the multi-band high resolution imaging  
to study each objects in the ACS band closer to the B rest-frame, in order
to minimize effects of morphological $K$-correction: the F435W filter is used 
in the range $0<z<0.2$, F606W in the range $0.2<z<0.55$,
F775W in the range $0.55<z<0.85$ and F850LP for $z>0.85$.
The z band is only a poor approximation
of the B-band rest-frame for objects with $z\gtrsim1.2$, approaching
towards to the U-band rest-frame. We will come back later to this problem.

\section{Data Analysis}\label{analysis}

The morphological analysis has been performed in three steps: $(a)$ a visual 
inspection, in order to assign each object to morphological classes based on 
the detected features (spiral arms, tails, double nuclei...); $(b)$ a 
surface-brightness profile analysis performed with GASPHOT and GALFIT, 
in order to quantify morphology and in particular to extract S\'ersic 
indices; $(c)$ a non-parametric analysis of the distribution of the galaxy 
light, using the measures of asymmetry, concentration and 
clumpiness (or smoothness) to separate different galaxy types.
In the following we will use only the results obtained by the visual 
inspection, reinforced by those from automatic procedures. However,
we are also interested in identifying a completely automatic procedure able
to segregate at least early- from late-type galaxies with a low level 
of interaction. This procedure will be useful for the incoming new very large 
and deep surveys like GOODS and COSMOS.

\subsection{Visual inspection and morphological classification}\label{visual}

The galaxies have been separated into five morphological classes 
(see Fig.\ref{mor_ex}):
1. Early-type galaxies (ellipticals and S0);
2. Peculiar early-types, that is either spheroidals objects 
showing some amount of isophotal asymmetry, or galaxy pairs in
which the main component is clearly elliptical;
3. Normal Spirals with regular disk structure and no 
evidence for luminous high-star formation regions; this kind of galaxies have
usually a luminous dominant bulge;
4. Perturbed Spirals, that is disk-dominated galaxies (with a central bulge)
showing an asymmetric structure due to interactions or to zones of strongly 
enhanced star formation;
5. Irregular galaxies, that is objects with a high degree of 
asymmetry showing no evidence for a disk component.

\begin{figure*}
\begin{center}
\includegraphics[width=\textwidth]{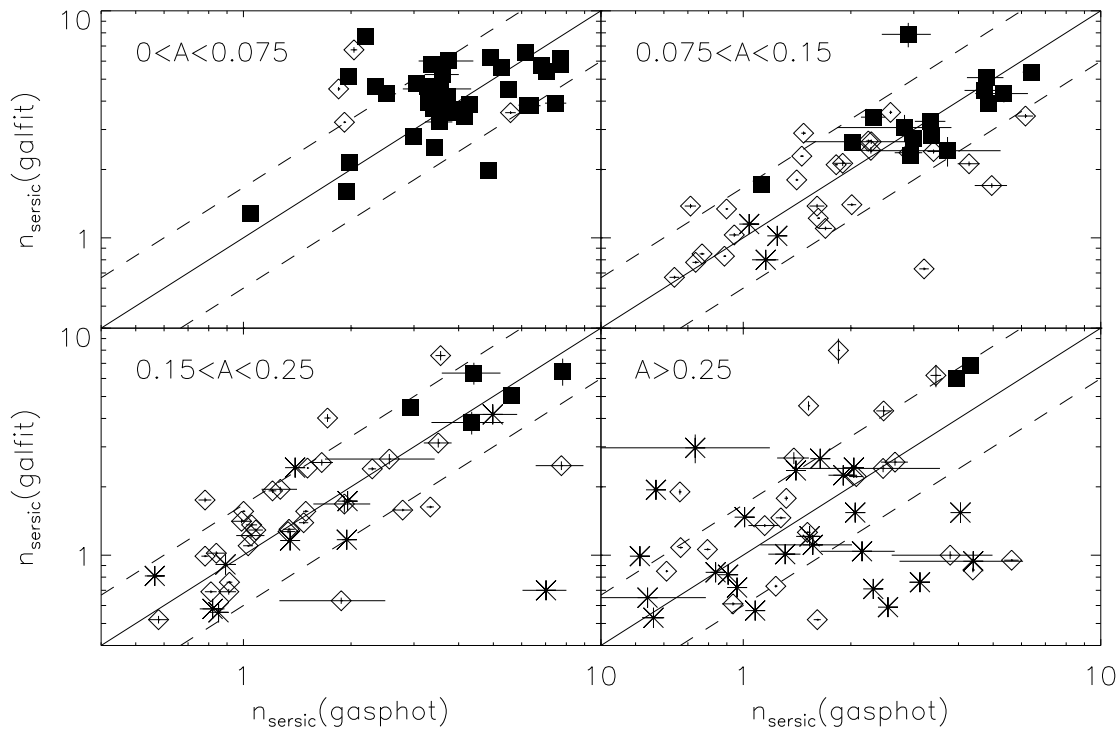}
\end{center}
\caption{
S\'ersic index retrieved by Gasphot versus those retrieved by Galfit  
for K20 galaxies in four bins of asymmetries (A). Dashed lines mark the 
region where $\Delta n/\bar{n}\le0.5$. Filled squares are visually classified
ellipticals/S0 galaxies (class 1 and 2), open diamonds are spirals (class 3 and 4) 
and crosses are irregulars.
} 
\label{sersic_gal_vs_gas} 
\end{figure*}

Figure \ref{mor_ex} illustrates some galaxies extracted from the 5 
morphological classes.
% the top-left panel shows a normal elliptical galaxy 
%representing the class 1; 
%the top-central panel is an example of isolated peculiar ellipticals with 
%signs for distortion of the isophotes; in the top-right panel there is a 
%couple of ellipticals, assigned to the class 2 of peculiar ellipticals; 
%bottom-left panel shows a normal spiral, dominated by the luminous central 
%bulge and showing a clear disk component with spiral arms; the bottom-central 
%panel contains a perturbed spiral, in which the bulge is less evident than 
%in previous case, but an evident asymmetric structure is present; in the 
%bottom-right panel there is a case of irregular galaxy.

The visual inspection has been performed independently by three authors,
PC, GR and GF, in order to minimize systematic trends. For the cases on 
which the classifications did not match, we have chosen the median 
morphological value.

Eventually, we have assigned 60 objects to class 1, 14 to class 2, 80 to class 
3, 48 to class 4 and  98 to class 5. 
%As it can be seen in Fig. \ref{frac_z}, the irregular morphology dominates at 
%all redshift.
The classes with peculiar disks and irregular galaxies are the most populated 
($\sim50\%$ of the galaxies fall in class 4 or 5).
It was not possible to systematically separate ellipticals from S0, given 
the reduced isophotal area covered by the sources.
Among the 46 galaxies belonging to the above mentioned clusters, 20 are normal 
ellipticals, 3 are peculiar ellipticals, 10 are normal spirals, 5 are peculiar 
spirals and 8 are irregulars.

\subsubsection{The effect of $K$-correction on the objects with $z>1.2$}\label{morkorr}

We analyze here the effects of $K$-correction on the morphological classification 
of objects having $z\gtrsim1.2$. The reddest ACS available band (the F850LP filter) 
does not match for these galaxies the B-band rest-frame, that is used for the 
morphological analysis of the remaining objects at lower z, but rather matches 
the U-band rest-frame. 
%The relative importance of star formation and old stellar 
%populations changes with the observation wavelengths: galaxies indeed 
%appear as later types at shorter wavelengths (Papovich 2003).

To this end we have selected a subsample of galaxies with $z\sim1$, and
we have compared their morphologies in the V 
(6060 \AA, corresponding to the U-band rest-frame) and $z$
(8500 \AA, corresponding to the B-band rest-frame) ACS bands.
In particular, 3 Elliptical/S0 galaxies ($1-3$ in 
Fig. \ref{kcorr}), 3 Normal Spirals ($4-6$ in Fig. \ref{kcorr}),
3 Perturbed Spirals ($7-9$ in Fig. \ref{kcorr}) and 3 Irregulars ($10-12$
are shown as examples in Fig. \ref{kcorr}).

\vspace{2mm}
\noindent{\it Ellipticals.} 
The galaxies in panel 1 and 3 have a clear elliptical 
morphology both in U as in B rest-frame. The galaxy in panel 2 represents
a border case: according to its B rest-frame image has been classified 
as S0 and assigned to class 2; it must be noted that this is the only case
in which the classifiers have been able to distinguish S0 from E or Sa 
morphologies. In the U rest-frame the S0 morphology is
roughly preserved, tending perhaps to resemble to a Sa Spiral. 

\vspace{2mm}
\noindent{\it Normal Spirals.} 
The galaxies in the panels 4, 5 and 6 have been classified
as normal spirals according to the presence in their B-band rest-frame
images of a central bulge on top of a spiral-dominated disk. It turns out that 
for the objects in panel 4 and 6 the morphology is preserved going from B to U 
rest-frame; in the remaining case the bulge becomes much less luminous in U and 
the star formation regions of the disk make the morphology looking 
more irregular.

\vspace{2mm}
\noindent{\it Perturbed Spirals.} 
The galaxies in the panels 7, 8 and 9 have been
classified as spirals according to the presence in their B-band rest-frame
of a disk structure with a central bulge (less luminous that in the above case).
The peculiarity depends on the presence of asymmetric structures (zones
of high star formation or interactions). Also in this case, the U rest-frame
luminosity of the central bulge decreases, making the
asymmetries more evident. In particular, in two cases (panels 7 and 8) the dominant
morphology becomes irregular, whereas perhaps in the latter case (panel 9)
the underlying disk structure and the spiral arms structure are preserved.

\vspace{2mm}
\noindent{\it Irregulars.}
In all the three cases the irregular morphology is preserved, as expected.
\vspace{2mm}

Observing the Fig. \ref{kcorr}, it turns out that the most important effect of 
the morphological $K$-correction concerns objects with a B rest-frame 
spiral-like morphology, which may become irregulars in the U rest-frame imaging. 
In order to quantify the percentage of mis-classified spirals at $z\gtrsim1.2$, 
we have performed the comparison between B and U rest-frame imaging over 
a sample of 27 disk galaxies at $z\sim1$.
We found that the morphology moved to irregular only in 5/27 cases (18\%).
Since we do not know if the $z\sim1$ galaxies 
are representative of the universe at higher $z$, this result suggests that at 
$z\gtrsim1.2$ the fraction of irregular galaxies could be slightly
over-estimated, as the fraction of spirals could be slightly under-estimated.

We have also performed the above analysis over a sample of 10 elliptical
galaxies at $z\sim1$, seeking cases similar to panel 2 of Fig. \ref{kcorr}.
We concluded that the the cited case is the only in which elliptical
morphology is not preserved moving from B to U rest-frame.
Then the number of elliptical galaxies at $z>1.2$ seems to be rather solid.

\subsection{Surface brightness profile analysis}
\subsubsection{The GASPHOT automatic analysis tool}

GASPHOT is a package for fully automatic surface photometry of galaxies, 
with a very low level of visual interaction, hence
particularly suitable for very large imaging datasets (Pignatelli 
et al. 2004). However, its validity for morphological classification is 
to be considered only in a statistical sense, rather than on a single 
object basis.

The tool uses a modified version of SExtractor (Bertin \& Arnouts 1996) 
to identify galaxies in the image and to extract isophotes. 
%It is possible to use an associated catalog containing the objects 
%of interest. 
Then it analyzes the profiles and derives the main photometric parameters 
of each identified object. 
The program determines the light growth-curves along the major and minor axes.
These are then fitted with a S\'ersic law ($\mu\propto r^{1/n}$), convolved with
the PSF, with five free parameters: the total magnitude $M_{tot}$, 
the half-luminosity radius in arcsec $r_e$, the S\'ersic index $n$, the axial 
ratio $b/a$ and the value of the local background.

This one-dimensional approach provides a more robust estimate than 
a more complex 2-D fit of the surface brightness image, which is more sensitive
to the presence of features like spiral arms, double nuclei or dust lanes. 
It is also less affected by instrumental artifacts in the real image,
particularly for very faint galaxies like our own.

Profiles with high values of the S\'ersic index imply an early-type 
morphology (the de Vaucouleurs profile has n=4), while low values usually indicate 
later-type morphologies (the exponential profile has n=1).
Unfortunately, many bulge-dominated objects have intermediate S\'ersic indices,
or even an exponential profile, so a residual degeneracy still remains in 
assigning the morphological class.

The situation gets more complicated when treating irregular galaxies and when 
dealing with blended objects. The isophotes flagged by SExtractor as 
blended are not used in the fit, in order to minimize distortions. This makes
the fitting procedure much more certain.
%By definition, GASPHOT is not useful for identifying irregular galaxies.
In this framework, irregular galaxies can only be identified by very large $\chi^2$
or failed fits. In these cases, however, the photometric parameters obtained are
obviously meaningless.

\subsubsection{The automatic analysis package GALFIT}

%\begin{table}%[!ht]
%\begin{center}
%\begin{tabular}{cccccc|c}
%\hline
%\hline
%&\multicolumn{2}{c}{\bf{unacceptable fits}}\\
% & GASPHOT & GALFIT\\
%$A<0.075$           & 1/64   & 4/64  \\
%$0.075<A<0.15$      & 11/57  & 15/57 \\
%$0.15<A<0.25$       & 6/63   & 8/63  \\
%$A>0.25$            & 28/116  & 37/116 \\
%%\vspace{3pt}\\
%\hline
%%\vspace{3pt}\\
%&\multicolumn{2}{c}{\textbf{$\#$ objects with}}\\
%&\textbf{$\Delta n/\langle n\rangle>0.5$}&\textbf{$\Delta r/\langle r\rangle>0.5$}\\
%$A<0.075$            & 11/47 & 18/47\\
%$0.075<A<0.15$       & 7/50  & 6/50\\
%$0.15<A<0.25$        & 10/52 & 11/52\\
%$A>0.25$             & 28/62 & 22/62\\
%\hline
%\hline
%\end{tabular}
%\end{center}
%\caption{
%In the upper panel we report the number of objects with an unacceptable 
%fit (that is a fit with $n_{Sersic}$=8 or $n_{Sersic}$=0.5 or $r_e$=0.7 
%pixels) for GASPHOT nd GALFIT packages as a function of the asymmetry bins. 
%In the lower panel we report among the objects with both GASPHOT and GALFIT 
%fits acceptable the number of objects having 
%$\Delta n/\langle n\rangle>0.5$ and $\Delta r/\langle r\rangle>0.5$
%for the same asymmetry ranges.
%}\label{gal_gas_t}
%\end{table}

GALFIT is an automatic tool to extract structural parameters from
galaxy images. At variance with GASPHOT, GALFIT fits the whole 2-D 
sky-projected light distribution. 
It combines together different parametric models (the Nuker law, the 
S\'ersic-de Vaucouleurs profile, the exponential disk, the Gaussian or 
Moffat functions) and allows multi-component fitting (useful to calculate 
e.g. Bulge/Total light ratios) and provides measures of the 
diskyness/boxyness of the examined galaxy.
If available, the PSF is used to convolve the model before fitting.
It is also possible to mask out of the fit peculiar regions (dust lanes, 
nearby companions, spiral arms, etc.) that the user wants to exclude.

\begin{figure}
\begin{center}
\includegraphics[width=\columnwidth]{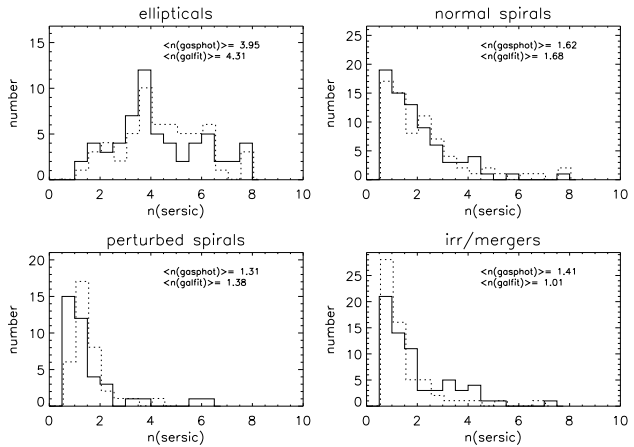}
\end{center}
\caption{
The distributions of S\'ersic indices retrieved by GASPHOT (continuous line) and
GALFIT (dotted line) for the four morphological classes. The mean value
of $n_{Sersic}$ for the two tools is reported in each panel.
}
\label{mclas_gal_vs_gas} 
\end{figure}

Even if GALFIT allows to reach a high level of detail in modeling
galaxy light, it needs substantially more interaction for each individual object
and it works well for bright galaxies with good sampling, 
rather than for the fainter ones closer to the sensitivity limit.

We have built a tool automatically running GALFIT over all 
galaxies of interest, making use of SExtractor to determine 
the required initial guess of the model parameters (magnitudes, 
scale radii, axial ratios and position angles).

\subsubsection{Comparison between GALFIT and GASPHOT}

Although GALFIT can in principle use multi-component models, in order to
compare with each other the results from GALFIT and GASPHOT, we have
forced GALFIT to the single component, S\'ersic model.

We must also first clear the sample from the objects for which the two tools 
were not able to produce a fit.
In fact, because of the numerical approximations involved and in order to 
avoid unrealistic run-away solutions, GASPHOT allows the fitting parameters 
to move during the fitting process in the region limited by $R_e >0.7$ pixels 
and $0.5 < n < 8$ only. Solutions whose best fit-parameters are beyond these 
boundaries are rejected, and the related galaxies are flagged as "failed fits".
GALFIT has no such limits: however, in order to perform a consistent comparison 
between the results of the two tools, we put the same limits  on the best-fit 
parameters of both tools and removed from the sample those for which one of the 
two tools was not able to find a solution.

In addition, due to the very low S/N ratio, GASPHOT and GALFIT were not
able to produce any fit for 16 and 14 galaxies, respectively.
Once we removed these objects, mainly irregular or high asymmetric objects,
the galaxies left in the sample are respectively 254 (GASPHOT) and 236 (GALFIT), 
of which 211 are in common.

In Fig. \ref{sersic_gal_vs_gas} we have reported the S\'ersic indices retrieved
by GASPHOT against those obtained by GALFIT, finding that 73\% of the sample is 
included in the region 
\mbox{$|n_{\rm GASPHOT}-n_{\rm GALFIT}|/\langle n\rangle\le 0.5$},
where $\langle n\rangle$ is the average of the two results, while 9\% have 
$\Delta n/\langle n\rangle \ge 1$.
It is worth stressing that, even though for toy galaxies the results
from GASPHOT and GALFIT have been found to differ less than 10\% from
one another (Pignatelli et al. 2004), the large scatter in Figure 3 is
not surprising when dealing with real featured galaxies, with light
distributions often asymmetric and far from being stick to the
models. In particular, due to the different fitting approaches adopted
(1D vs. 2D), we expect that GASPHOT and GALFIT provide rather
different results at increasing asymmetry. Thus, we split the Figure~3
in different bins of the asymmetry parameter A.
Actually, apart from the first asymmetry bin, the scatter around the 1:1 relation
increases with the asymmetry of the galaxies analyzed, and a similar trend
is found for the number of failed fits as a function of the asymmetry:
more asymmetric galaxies are obviously harder to model with a spherically symmetric 
S\'ersic law.
The large scatter observed in the first bin of A is due to the
predominance of early-type galaxies (high values of S\'ersic index) and
to the fact that the uncertainty of the retrieved values of $n$
intrinsically increases with the S\'ersic index itself (Pignatelli et
al. 2004).

\begin{figure*}
\begin{center}
\includegraphics[width=\textwidth]{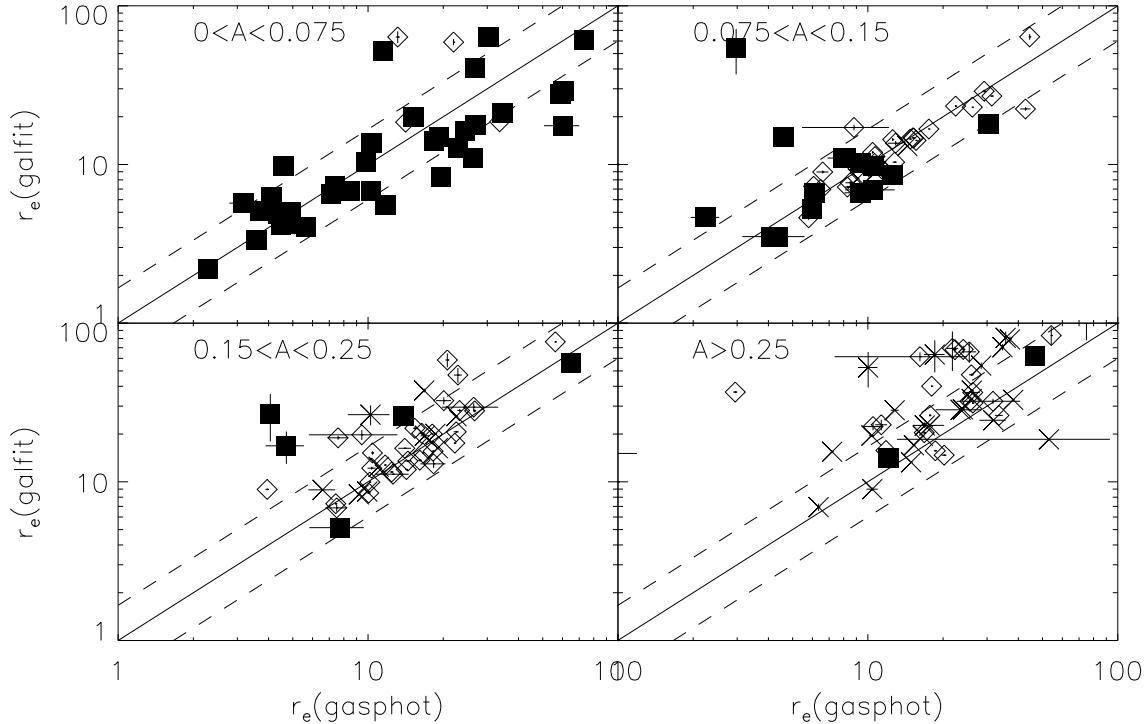}
\end{center}
\caption{
Scale radius retrieved by GASPHOT versus those retrieved by GALFIT for 
K20 galaxies in four bins of asymmetry. Dashed lines mark the region where 
$\Delta r_e/\bar{r}_e\le0.5$. Filled squares are 
ellipticals/S0 galaxies (class 1 and 2), open diamonds are spirals 
(class 3 and 4) and crosses are irregulars.
}
\label{re_gal_vs_gas} 
\end{figure*}

A similar trend is visible in Fig. \ref{re_gal_vs_gas}, where we show the 
comparison between the optical radii retrieved by the two tools. Even now 73\% 
of the sample is comprised in the region
 $|r_{\rm GASPHOT}-r_{\rm GALFIT}|/\langle r\rangle\le 0.5$
In this case, however, the scatter around the 1:1 relation is not symmetric:
there is a large amount of galaxies for which the GALFIT scale radius is 
considerably larger than that obtained by GASPHOT. A similar effect was also 
found by Pignatelli et al. (2004) in a systematic comparison between the two tools.

We checked visually the objects for which this difference was more noticeable. 
We found that, for compact objects with a small isophotal radius and high 
S\'ersic index, GALFIT can sometimes recover unrealistic optical radii, often 
many times larger than the isophotal area itself. 

Finally, we want to investigate further whether a criterion based on the S\'ersic
index could be able to separate early- from late-type galaxies. Ravindranath 
et al. (2004) for example classified as bulge-dominated and disk-dominated galaxies
objects with $n_{Sersic}\le2$ and $n_{Sersic} > 2$, respectively.

In Fig. \ref{mclas_gal_vs_gas} the distributions of the S\'ersic indices obtained
by GASPHOT and GALFIT for the five morphological classes are reported
(normal E's/S0 and perturbed ellipticals are plotted together).
It is worth noticing that, while the results of the two tools for individual objects
may differ, the statistical distributions are quite similar.

The number of elliptical/S0 galaxies (morphological class 1 and 2) which have
S\'ersic indices lower than 2 are only 6 and 3 according to GASPHOT and
GALFIT respectively.
Instead, the contamination of late type galaxies with S\'ersic indices greater
than 2 is larger (56/254 and 50/237, according to GASPHOT and GALFIT respectively).
From Figure 4 we must conclude that the S\'ersic index alone just
provides a broad, not univocal indication of the morphological type.
Since this index is a measure of the concentration of light, this is
in agreement with the results of Abraham et al. (1996), who found
that, besides the Concentration, at least one more parameter is needed
to classify galaxies.

\subsection{The CAS parameter set}

The non-parametric methods for morphological estimates are those which do not need
to assume a parametrized analytic function to model the galaxy light distribution.
They constitute an important complement to the problem of quantitative morphology 
(Abraham et al. 1996, Conselice et al. 2000, Conselice 2003).
We use the classical Concentration, Asymmetry and clumpineSs (or smoothness) 
parameters. 
The Concentration correlates with the S\'ersic index: high Concentration 
values correspond to early-type morphology, while lower values are suggestive of 
a disk-dominated or irregular galaxy.
The Asymmetry can distinguish irregular galaxies or perturbed spirals from 
relaxed systems as E/S0 and normal spirals.
The Clumpiness quantifies the degree of structure on small scales, and roughly
correlates with the rate of star formation.

\begin{figure*}
\begin{center}
\includegraphics[width=\textwidth]{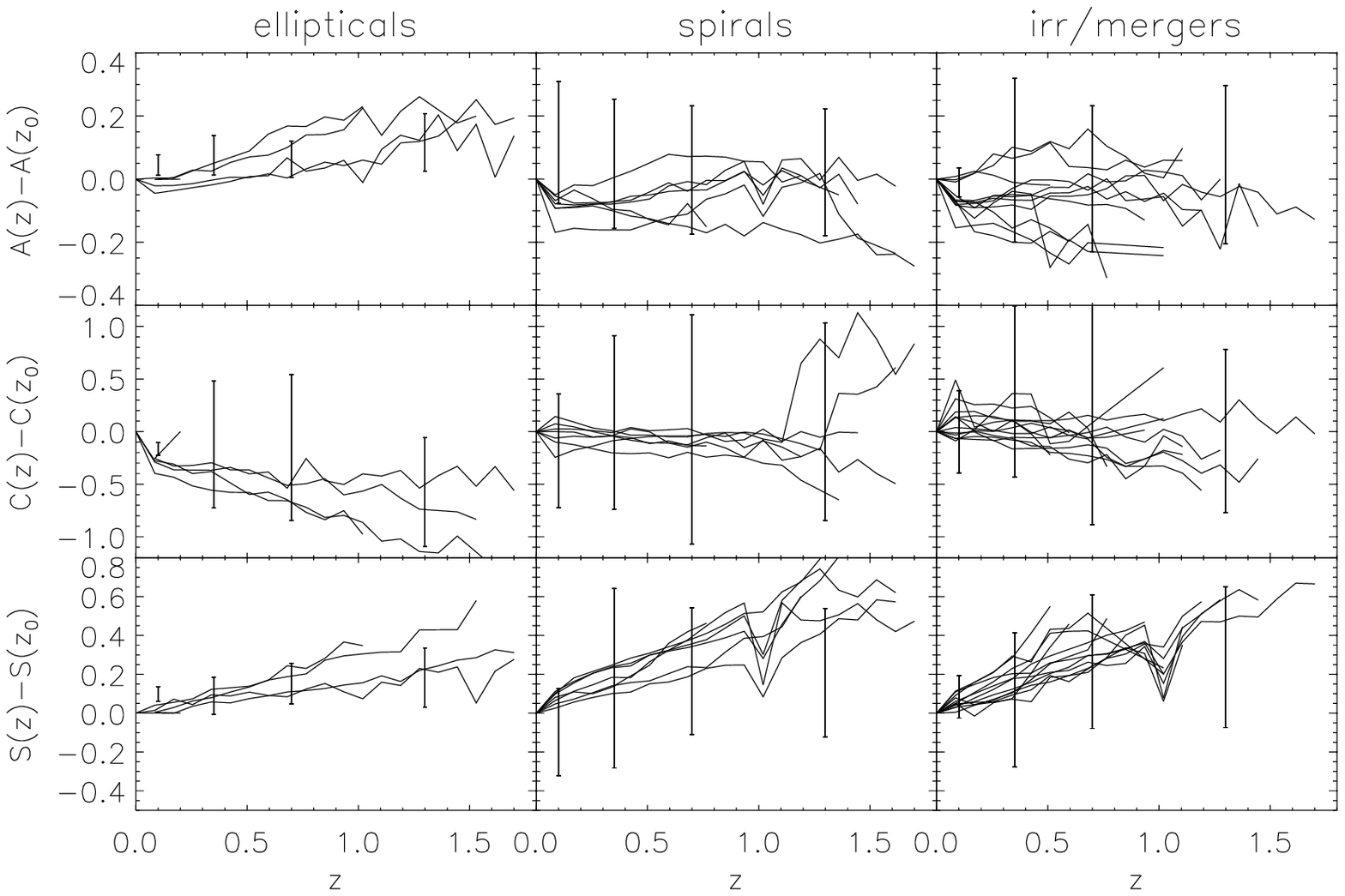}
\end{center}
\caption{
Evolution of the CAS parameters with $z$ for a sample of artificially
redshifted galaxies. 4 ellipticals, 7 spirals and 11 irregular galaxies
are simulated at increasing redshift to isolate systematic effects. The
resulting parameters for the entire sample in the four redshift bin is
shown for comparison (vertical bars).
}
\label{cas_z} 
\end{figure*}

\subsubsection{Definitions}

The operational definitions of the three parameters may differ from author to
author. We refer to the Conselice et al. (2003) definitions, but we modified 
them in order to make simpler and faster the computations.
The Concentration is a logarithmic ratio of the apertures containing 
80\% and 20\% of the total flux:
\begin{equation}\label{def_conc}
C=5\log\left(\frac{r_{80}}{r_{20}}\right).
\end{equation}
operatively, we interpolate the flux growth curve obtained using 12 aperture 
radii within which SExtractor calculates the flux.

The Asymmetry measures how much the galaxy light is symmetric with respect to
a rotation of 180$^{\circ}$ around the galaxy centroid. So the Asymmetry is 
qualitatively the residual of the difference between the original $I_0$ 
and the rotated $I_{180}$ galaxy image:
\begin{equation}\label{def_asym}
A=\frac{\sum\mid I_0-I_{180}\mid}{2\sum\mid I_0\mid} .
\end{equation}
In order to reduce the influence of the background noise, we subtract from 
the galaxy image the background given by SExtractor, and we remove those 
pixels having values lower than 1.5 times the $rms$ of the background. The results 
strongly depend on the adopted galaxy centroid. We solve this problem by 
computing the Asymmetry on a grid of $100\times100$ points around the 
center calculated by SExtractor. The distance between contiguous points
in this artificial grid is 0.4 pixels, so the Asymmetry is computed for 
points up to $\pm20$ pixels far from the initial center. The Asymmetry 
is defined as the minimum of the values computed on the whole grid. 

The Clumpiness measures the fraction of galaxy light lying in the high
spatial frequencies. It is computed by subtracting from the original an 
image smoothed by a box of a given size:
\begin{equation}\label{def_clum}
%S=\sum_{x,y}\frac{\left(I_0(x,y)-I_{\sigma}(x,y)\right)}{I_0(x,y)}
S=\sum\frac{\left(I_0-I_{\sigma}\right)}{I_0}
\end{equation}
where $I_0$ is the original image and $I_{\sigma}$ is that convolved with 
a box of scale $\sigma$.
The value of the retrieved clumpiness depends strongly by the scale used 
for the convolution. We use $\sigma=2/3r_{SEx}$ and $\sigma=r_{SEx}$ 
(where $r_{SEx}$ is the FLUX\_RADIUS parameter given by SExtractor) 
respectively for galaxies with $r_{SEx}$ lower and greater than 5 pixels.
Again, in order to reduce the influence of the background noise we remove
those pixels having values lower than 2 times the $rms$ of the background.

To automatically compute the CAS parameters for all the galaxies in our 
image we have used specifically a designed IDL tool which uses 
SExtractor to produce the flux growth curves of galaxies, then IDL calculates 
the concentration indices, subtracts the background and filter noisy pixels. 
Finally a postage cut for each galaxy of interest is produced and these postages 
are suitably rotated and convolved to calculate Asymmetry and Clumpiness. 

We compute the CAS parameters in the ACS bands roughly corresponding 
to the B-band rest frame for each galaxy. 

\begin{figure*}
\begin{center}
\includegraphics[angle=90,width=\textwidth]{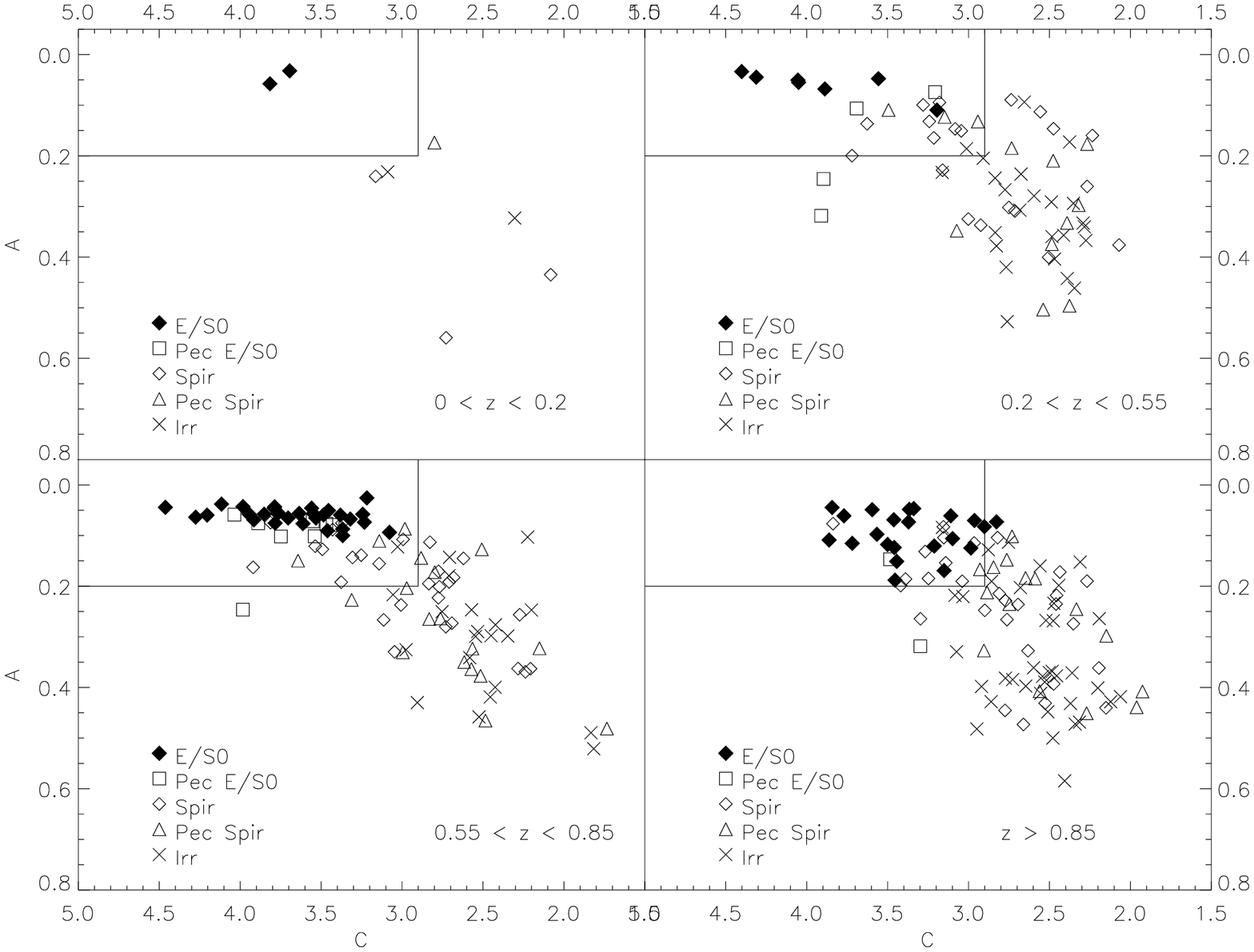}
\end{center}
\caption{
Evolution of the distribution of galaxies in the Concentration-Asymmetry 
plane in four redshift bins. The parameters are retrieved in the ACS band
nearest to the B-rest frame: galaxies with $z<0.2$ are analyzed in the
F435W filter, objects with $0.2<z<0.55$ in the F606W filter, objects
with $0.55<z<0.85$ in the F775W filter and those with $z>0.85$ in the
F850LP filter. The rectangle marks the region populated by ellipticals/S0 
galaxies.
}
\label{ca} 
\end{figure*}

\begin{figure*}
\begin{center}
\includegraphics[angle=90,width=\textwidth]{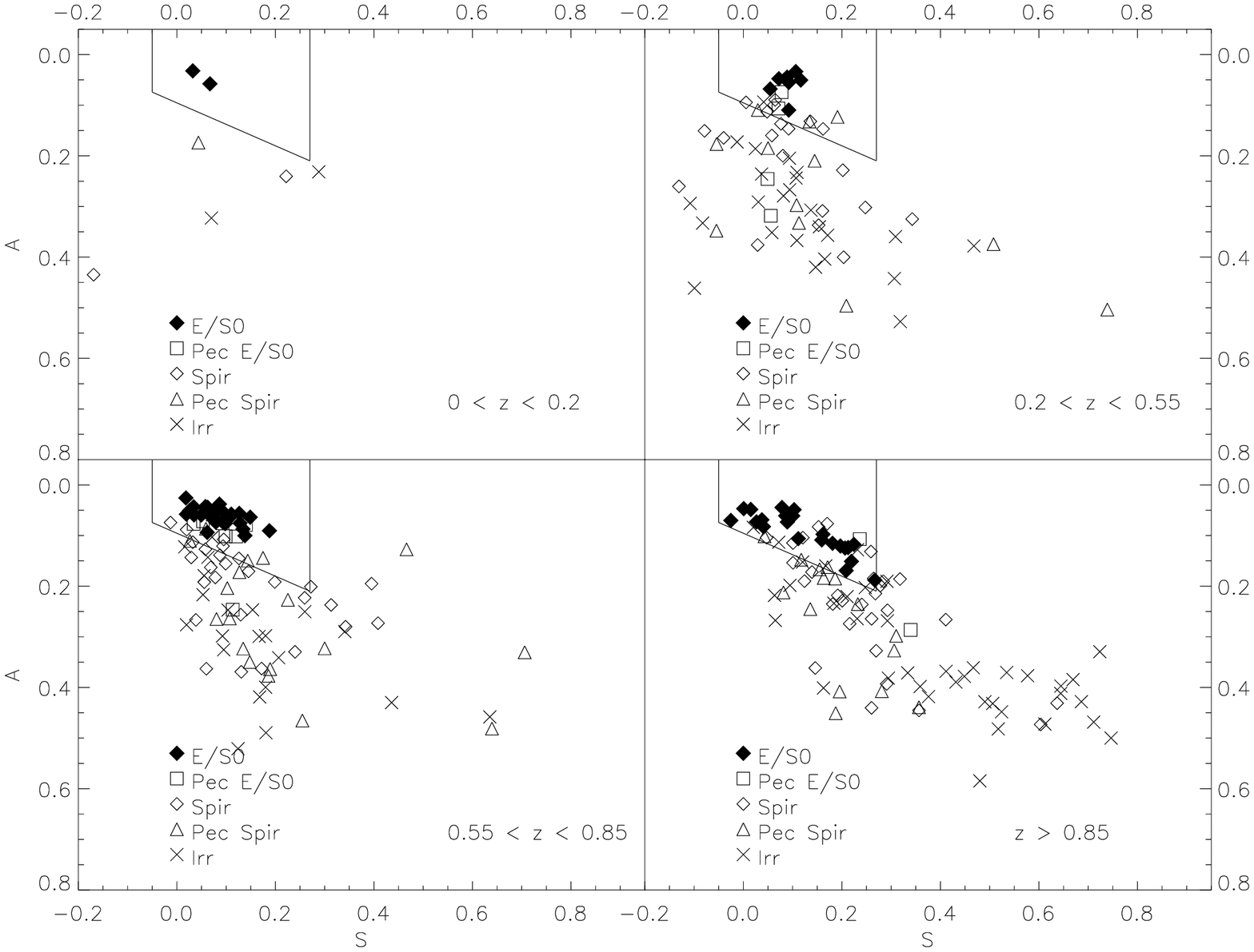}
\end{center}
\caption{
Evolution of the distribution of galaxies in the clumpineSs-Asymmetry plane
in four redshift bins. The parameters are retrieved in the ACS band
nearest to the B-rest frame: galaxies with $z<0.2$ are analyzed in the
F435W filter, objects with $0.2<z<0.55$ in the F606W filter, objects
with $0.55<z<0.85$ in the F775W filter and those with $z>0.85$ in the
F850LP filter. The lines marks the region typically 
populated by ellipticals/S0 galaxies.
}
\label{sa} 
\end{figure*}

\begin{figure*}
\begin{center}
\includegraphics[angle=90,width=\textwidth]{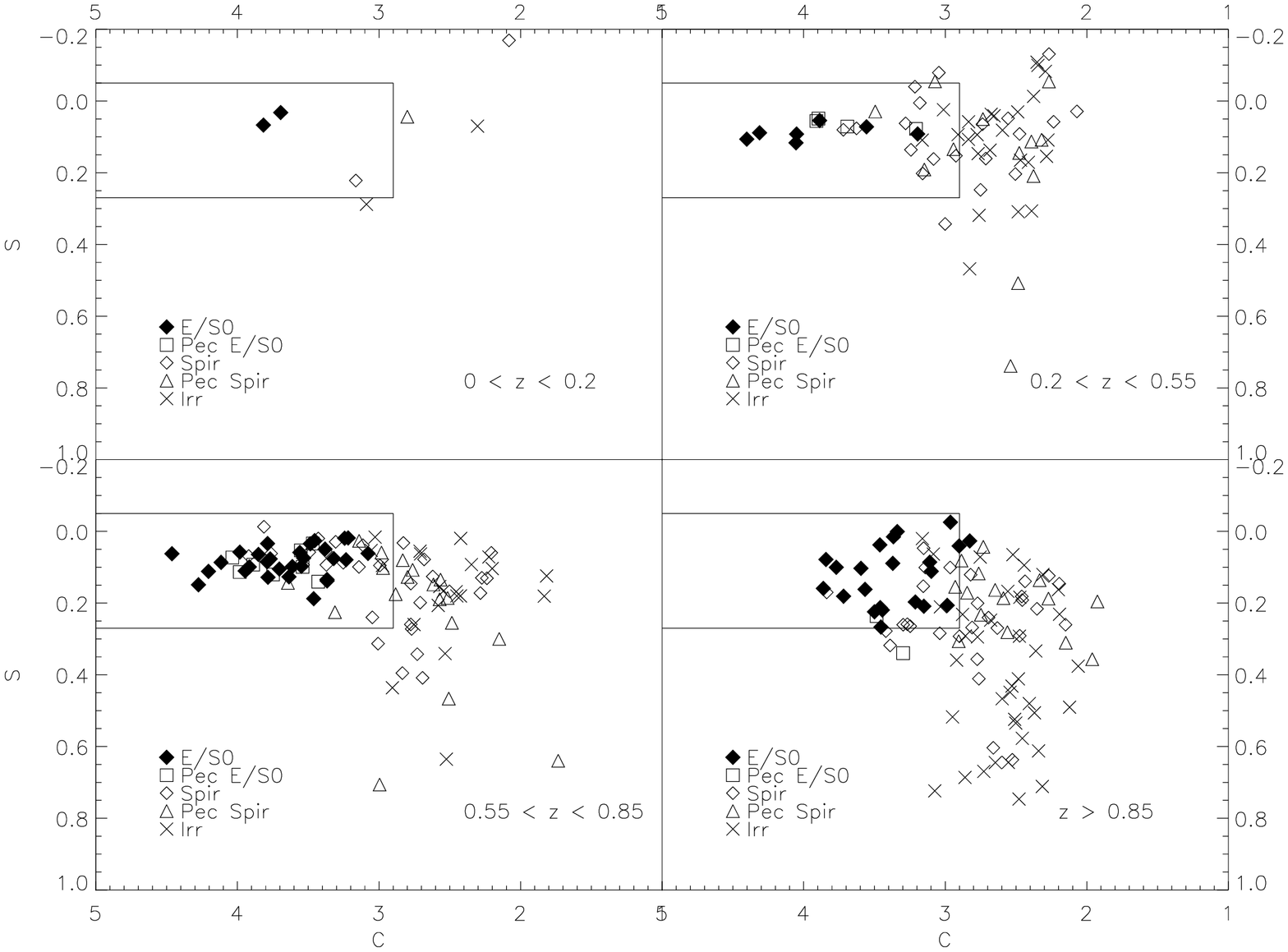}
\end{center}
\caption{
Evolution of the distribution of galaxies in the Concentration-clumpineSs plane
in four redshift bins. The parameters are retrieved in the ACS band
nearest to the B-rest frame: galaxies with $z<0.2$ are analyzed in the
F435W filter, objects with $0.2<z<0.55$ in the F606W filter, objects
with $0.55<z<0.85$ in the F775W filter and those with $z>0.85$ in the
F850LP filter. The rectangle marks the region typically 
populated by ellipticals/S0 galaxies.
}
\label{cs} 
\end{figure*}

%\subsubsection{Random Errors}

%\begin{table*}%[!ht]
%\begin{center}
%\begin{tabular}{cccccccccccccccc}
%\hline
%&& & Ellipticals & & & & & Spirals & & & & & Irregulars & &\\
%&& C & A & S & & & C & A & S & & & C & A & S\\
%\hline
%$z\sim0.2$ && 0.01 & 0.01 & 0.01 &&& 0.01 & 0.01 & 0.02 &&& 0.01 & 0.01 & 0.01\\
%$z\sim0.6$ && 0.02 & 0.01 & 0.02 &&& 0.01 & 0.02 & 0.02 &&& 0.02 & 0.02 & 0.01\\
%$z\sim1.2$ && 0.02 & 0.02 & 0.02 &&& 0.02 & 0.02 & 0.02 &&& 0.03 & 0.03 & 0.02 \\
%\hline
%\end{tabular}
%\end{center}
%\caption{
%Random errors of the CAS parameters for ellipticals, spirals and irregulars at 3
%different redshifts.
%}\label{cas_err}
%\end{table*}

%To estimate random error associated to CAS parameters we performed CAS analysys
%over a set of galaxies with different suimulated noise. 

\subsubsection{Simulation of CAS parameters for high-z galaxies}\label{High-z CAS parameters simulation}

The poor sampling of high z galaxies is likely to affect the measures of the CAS 
parameters. To check and quantify this point, we have 
simulated the effects of moving local galaxies with known CAS values 
to higher redshifts.
We have used as test objects all the galaxies in the sample 
having \mbox{$z<0.3$}. Following Conselice (2003), the angular size of galaxies 
is reduced by a rebinning factor $b$ given by the ratios of the angular 
diameter distances:
\begin{equation}\label{rebinning}
b=\frac{d_A(z_2)}{d_A(z_1)}
\end{equation}
where $z_1$ and $z_2$ are respectively the initial and final redshifts.
Then, pixel fluxes are reduced by the cosmological $(1+z)^4$ factor and 
are K-corrected.
Operatively, we first filter out the noise of the test galaxies with a simple
smoothing algorithm, and, after rebinning, changing the angular sizes and 
correcting the pixel fluxes, we convolved the images with a model of the PSF. 
Finally the noise is re-added and the CAS parameters are computed.

In Fig. \ref{cas_z} we report the results of these simulations for 
4 ellipticals, 7 spirals (including normal and perturbed spirals) 
and 11 irregulars galaxies. For comparison, the ranges of values that 
we measured for the 300 galaxies in our sample are reported as vertical 
bars. Although the number of simulated objects is not large, 
we note that the Asymmetry A does not depend on $z$ for spirals and irregulars, 
apart from an increase of the scatter at increasing redshift. 
On the other hand, a small increment of the mean 
value with z is present for the ellipticals. 

There is a moderate dependence on redshift in the estimate of Concentration 
C for the ellipticals: values at $z>0.5$ are systematically lower than local values.
Again, the scatter increases with the redshift for all classes.

As for the Clumpiness S, a bias appears to be introduced for all
morphological classes, since the mean value increases artificially with 
redshift, while the scatter remains moderate.

Spirals and irregulars display a large spread in the values of the 
Concentration and Clumpiness which are not reproduced by our simulations, 
likely due to the small size of the local reference sample.

Accortding to the measures of non-parametric indicators on real galaxies,
up to redshift $z\simeq 1.5$, we have highlighted some biases in the CAS 
structure of galaxies, introduced by instrumental effects.

In conclusion, our simulations suggest that CAS 
parameters provide an effective tool to analyse and discriminate galaxy 
morphologies in the z-interval of the K20 sample.

\subsubsection{Results of the CAS analysis}

Figures \ref{ca}, \ref{sa} and \ref{cs} show the distribution of 
our galaxies sample in the CAS space in various redshift bins. The different 
symbols refer to our visual morphological classification as discussed 
in Sect. \ref{visual}.

Objects in the morphological class 1 (ellipticals and S0s) separate 
significantly from the other classes. We have identified, independently 
from the redshift bin, the domains populated mainly by early-type 
galaxies, which are bounded by the continuous lines in the figures. 
These boundaries are given by the following conditions: 
\begin{equation}\label{earlycas}
A<0.2, \ \ \ \ \ \ C>2.9,  \ \ \ \ \ \ \ 0.05<S<0.27 \ \ \ \ \ A<0.45\times S+0.085.
\end{equation}
We stress that these boundaries are independent of the galaxy redshift.
Only 1 elliptical galaxy lies outside from these regions (it is in fact 
a very small galaxy with an uncertain classification). 

There are however 20 late-type galaxies out of 226 (9\%), as judged from 
visual inspection, which fall within the early-type galaxy CAS domain, 
and so would be misclassified by such criterion. It is worth to note 
that using different boundaries for each redshift bin reduces the 
contamination of late-type in the ellipticals domain of a small amount.
This contamination is due to the following objects: 8 galaxies 
dominated by a luminous and symmetric bulge, but with evidence for a 
faint disk; 3 ellipticals with a small companion in interaction; 
5 very compact objects, for which CAS estimate is uncertain;
4 further symmetric and concentrated galaxies, but clearly patchy, for
which the measure of the clumpiness essentially fails because of the small 
area and high ellipticity.

The datapoints corresponding to elliptical galaxies within the boundaries 
defined by eq.(\ref{earlycas}) show a scatter which sensibly depends 
on redshift: while it stays small up to $z=0.85$, it tends to explode above this
limit.

On the other hand, galaxies belonging to class 3, 4 and 5 
do not show a tendency to segregate in the CAS plots, but they occupy 
almost the same regions. 

In conclusion, the application of the CAS analysis to the K20 galaxies 
suggests this technique to be very efficient in disentangling early-type 
from late-type and irregular galaxies, with few percent of contamination. 
On the contrary, we failed in identifying algorithms able to 
resolve the different classes contributing to the late-type category.

\subsection{The Clumpiness-Asymmetry Relation for Galaxies}

\begin{figure*}
\begin{center}
\includegraphics[angle=90,width=\textwidth]{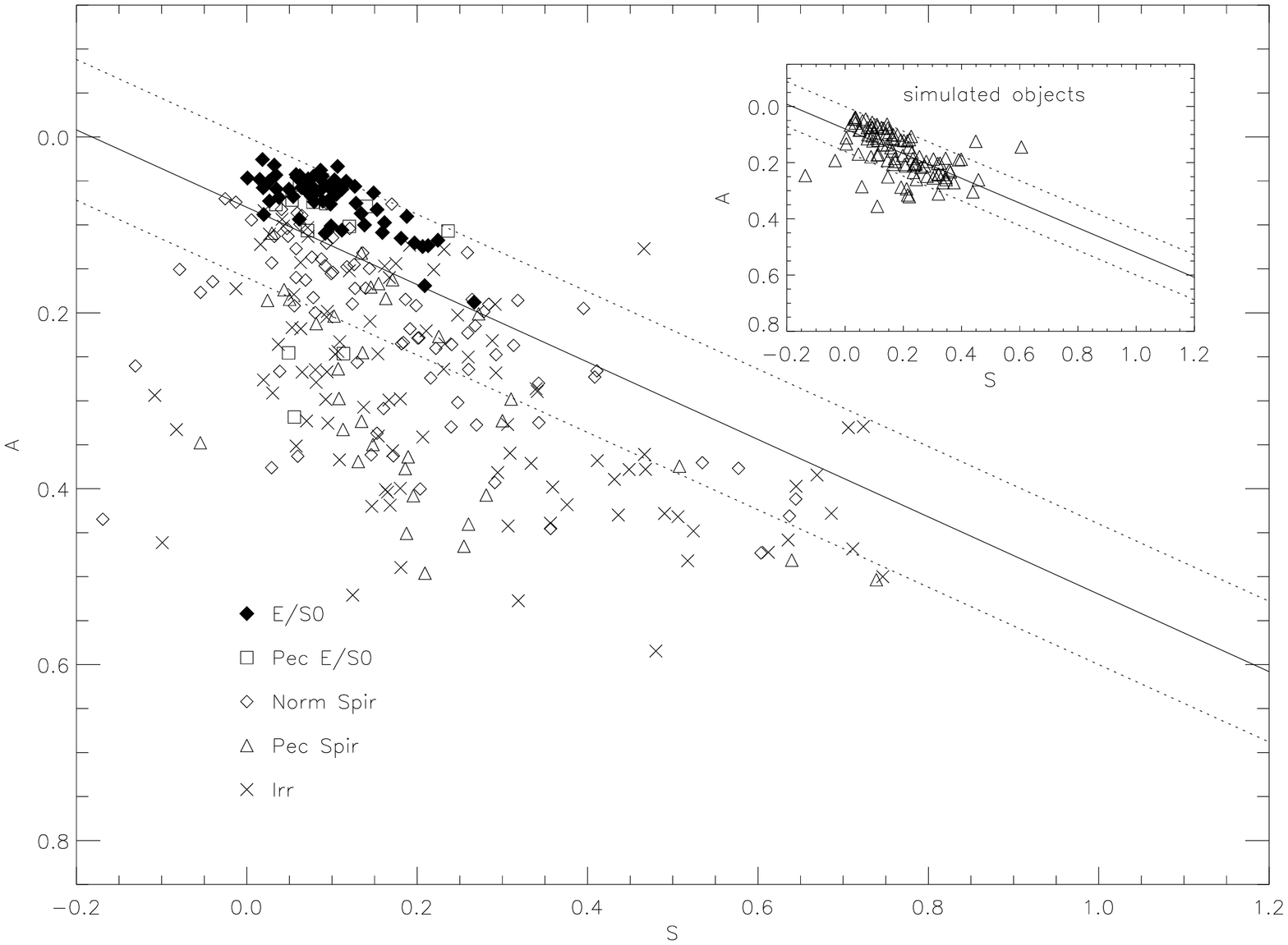}
\end{center}
\caption{
The distribution of the sample galaxies in the S-A plane. The strip between
dotted lines shows the region containing normal galaxies, calibrated using 
the normal galaxies of the sample simulated at various redshifts, shown in the
top-right panel.
}
\label{sa_mm}
\end{figure*}

Conselice (2003) noticed that the Asymmetry and Clumpiness parameters 
for normal galaxies in the local universe are correlated, populating 
a strip in the S-A plane. 
On the other hand, merging systems or irregular galaxies have typically 
the same clumpiness of the interacting components but higher asymmetry, 
hence they deviate from the relation. 
Fig. \ref{sa_mm} confirms this trend in our sample: the two parameters 
are clearly correlated for normal ellipticals and normal spirals. 
Because of the different operative definitions of the CAS parameters 
with respect to those of Conselice (2003), we needed to recalibrate the
relation. 
To this end we used the simulations described in Sect. 
\ref{High-z CAS parameters simulation} using data on 4 ellipticals and 
3 normal spirals and changing their redshift: our simulated A-S relation 
for normal (E/Sp) galaxies is reported in the inset of Fig. \ref{sa_mm}, 
together with the strip containing the 90\% of the data points. The 
best-fit relation and its 90\% boundaries are:
\begin{equation}
A=(0.44\pm0.10)S+(0.08 \pm 0.08).
\end{equation}
Fig. \ref{sa_mm} reports the distribution of the galaxies in the S-A plane
and the strip in which we expect to find only normal galaxies (defined as the
strip that contain 90\% of the simulated objects). 
The plot confirms that the large majority of visually inspected normal 
spirals and ellipticals fall indeed within the boundaries. 
Outside them only irregulars, peculiar spirals and some normal spiral
are found.

However, this criterion appears of limited usefulness to disentangle 
normal spirals from irregulars because a large number of visually 
classified irregulars also fall in the normal galaxy region.

%\subsection{Automatic Identification of Mergers}

%Due to the intrinsic scatter in retrieving the CAS parameters (see figure 
%\ref{cas_z}) it is not possible indeed locate a precise part of the plane
%populated by major mergers, but the number of this kind of objects 
%increases moving perpendicularly to the best-fit relation towards bottom-left 
%part of the S-A plane.

\section{The Redshift Distributions of the Morphological Classes}\label{frac}

\begin{figure}
\begin{center}
\includegraphics[width=\columnwidth]{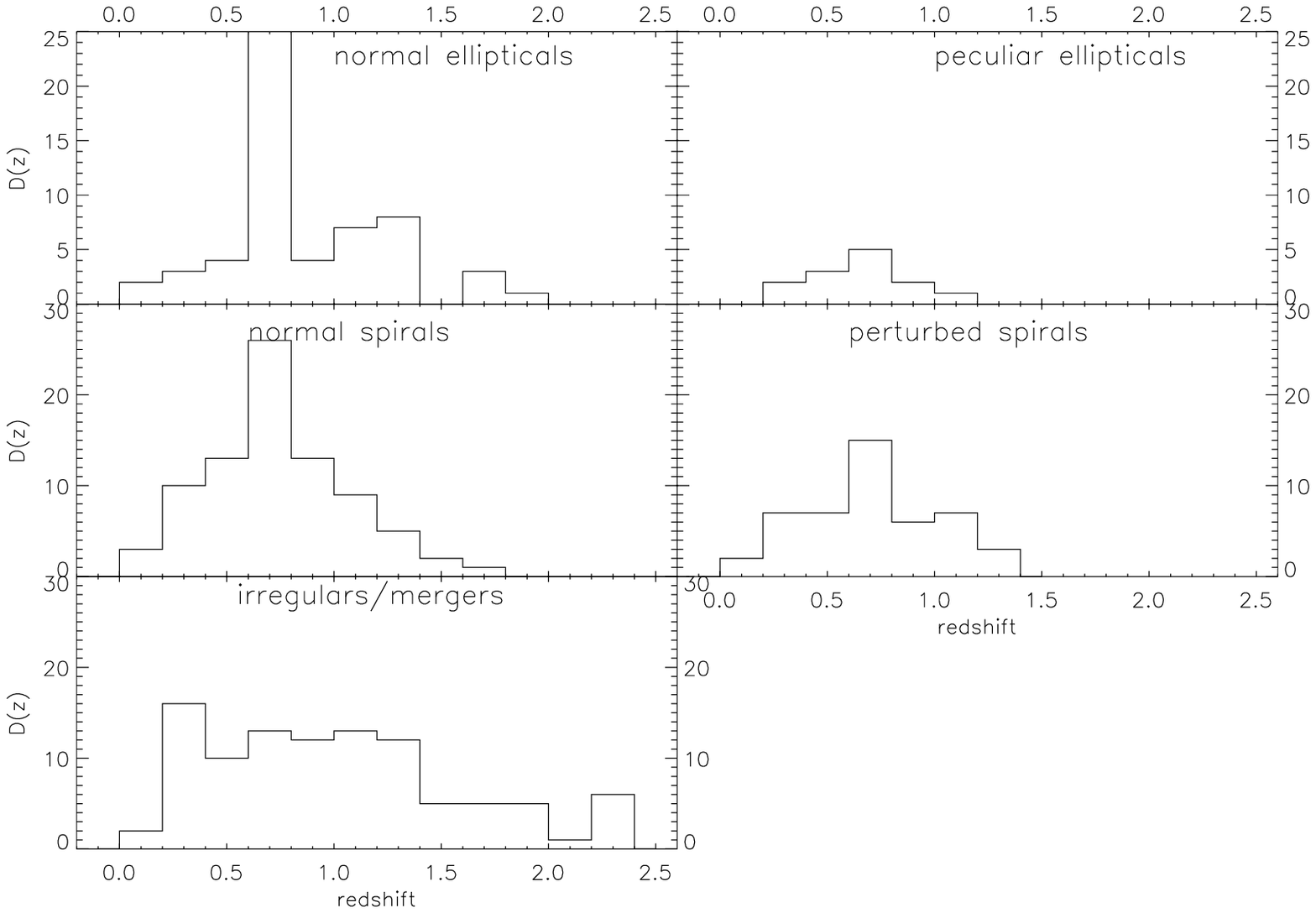}
\end{center}
\caption{
Redshift distributions for the five morphological classes.
}
\label{dz} 
\end{figure}

In Fig. \ref{dz} the redshift distribution for all the five morphological 
classes are reported. It should be noted the excess of early type galaxies
and spirals at redshift $z\sim0.75$, signature of the cluster of 
galaxies at $\bar{z}=0.737$. The highest redshift elliptical is is at $z=1.903$.
No disk galaxies lie at $z>1.8$. Irregular galaxies instead are numerically
relevant from $z=0.2$ up to $z=2.5$.

\begin{figure}%[!ht]
\begin{center}
\includegraphics[width=\columnwidth]{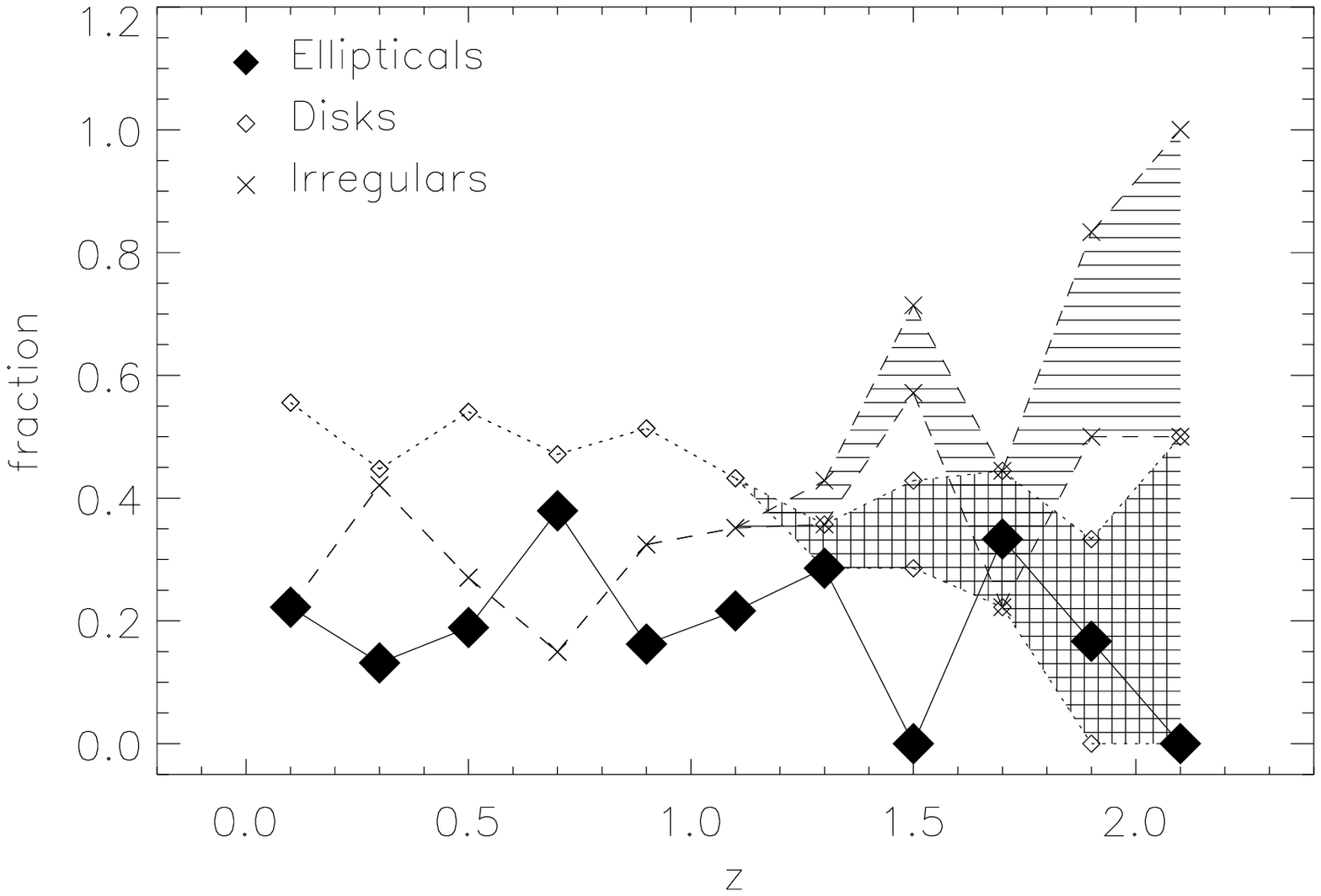}
\end{center}
\caption{
Morphological fractions as a function of the redshift. The results for
$z\gtrsim1.2$ are likely to be affected by morphological $K$-correction,
given that at those redshifts z-band does not match more the B rest-frame, 
but the U rest-frame. We showed in Sect. \ref{morkorr} that 
the $\sim20\%$ of the galaxies with $z\sim1$ classified as spirals in their 
B rest-frame appear as irregulars in their U rest-frame. The shaded regions
show the confidence interval for irregulars and spirals galaxies. The datapoint
at z=0.7 is affected by the presence of the two clusters.
}
\label{frac_z} 
\end{figure}

Figure \ref{frac_z} shows the evolution of the fraction of morphological 
classes with the redshift. For clarity we used here 
only three main classes: early types (including type 1 and 2), disks (type 
3 and 4) and irregular/peculiar (type 5). The shaded regions account for the
effects of morphological $K$-correction, as described in Sect. \ref{morkorr}: the
upper limit for the irregulars (and the lower for the spirals) corresponds to
the (unrealistic) case in which the $K$-correction does not affect the
classification of objects with $z\gtrsim1.2$; the lower limit for the irregulars
(and the upper for the spirals) shows instead the case in which $\sim20\%$ of the
morphologically classified irregulars at $z\gtrsim1.2$ are spirals.

The more evident result is the fast growth of the fraction of irregulars above 
$z\sim0.8$, where they are the dominant population of the sample, being at 
$z\gtrsim1.5$ more than the 60\% of the entire population.
The Elliptical fraction remains near to 20\% up to $z=1.5$ (beyond this redshift
the statistic is too poor), except for the point at $z\sim0.75$, where the
fraction is higher due to presence of the two clusters. The fraction of disk galaxies 
remains rather constant to the 50\% up to $z\sim1$, then it decreases rapidly. 
%The increasing number of irregular/peculiar galaxies with the redshift,
%together with the decrease of the ellipticals/disks implies that 
%some mechanism transforms a certain fraction of these irregular into the
%local spirals and ellipticals.

Conselice et al. (2004), studying the B-band rest-frame morphology of an 
I-selected sample of galaxies in the HDFs up to z=3, reach similar 
conclusions (see Fig. 9 therein): they also found a rapid decrease of 
ellipticals and spirals above $z\sim 1$ and a contemporary increase in the 
number of peculiar galaxies.
Our result however seems to be more robust at least up to $z\sim2$: on the one 
hand, the K-band selection ensures a better mass sampling than the optical 
selection, and on the other hand, the larger area minimizes effects of cosmic 
variance. The accurate investigation of the effects of morphological 
$K$-correction conducted in previous sections avoids biases in the morphological 
fraction up to $z\sim2$. 

%We stress that in this work we have conducted an accurate check of the effects
%of the morphological $K$-correction, making the results robust up to $z\sim2$.
%We stress our results are likely not suffering from the morphological 
%$K$-correction up to $z\sim1.2$, thanks to the B-band rest-frame analysis 
%we have conducted, and a very accurate evaluation of the effects at higher
%redshifts has been conducted.

\begin{figure*}%[!ht]
\begin{center}
\includegraphics[width=\textwidth]{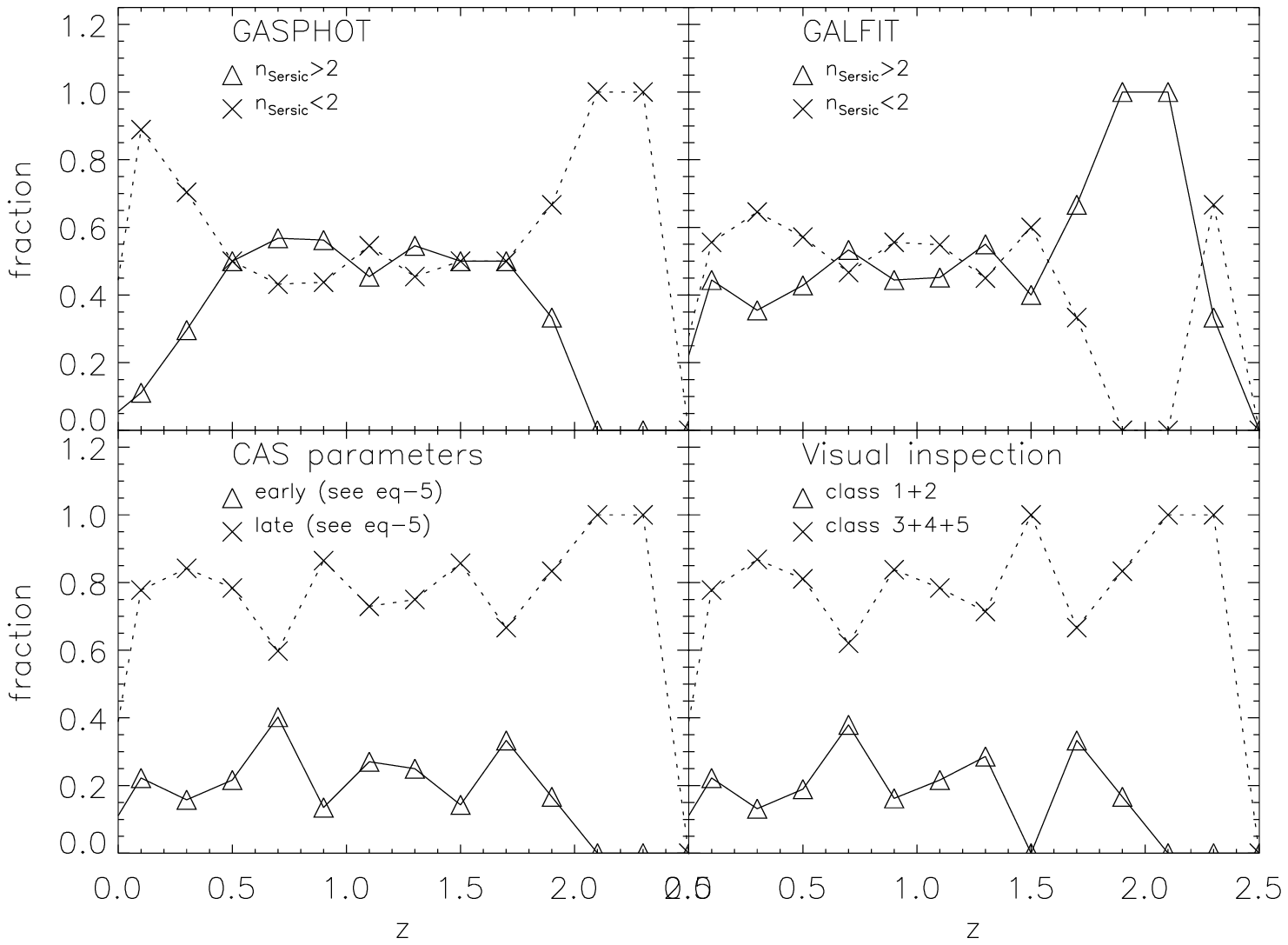}
\end{center}
\caption{
Morphological fractions for early- and late-type galaxies 
as a function of the redshift obtained by the different 
methods proposed in this paper: in the top two panels the criterion
based on $n_{Sersic}\gtrless2$ is used for GASPHOT and GALFIT;
in the bottom-left panel early-galaxies are selected according to
Eq. \ref{earlycas}; in the bottom-right panel the visual classification
is used.
}
\label{frac4_z} 
\end{figure*}

In Fig. \ref{frac4_z} a comparison of the morphological fractions 
(early- and late-type only) obtained with various methods explored in this 
paper is proposed. In the top two panels the fractions of objects with 
$n_{Sersic}\gtrless2$ are shown against the redshift according to GASPHOT 
and GALFIT. In the bottom-left panel the early-type galaxies are selected 
according to their CAS parameters (see Eq. \ref{earlycas}). Finally, in 
the bottom-right panel the results by the visual inspection are reported.
Among the automatic methods, only the CAS parameters one is able to reproduce
the results by the visual inspection. Among the $\sim250$ objects for which
GASPHOT and GALFIT give acceptable fits, $\sim50\%$ have $n_{Sersic}>2$. The
late-type fraction obtained with the $n_{Sersic}$ parameter only suffers of 
two effects: the large number of irregular high asymmetric objects for which 
GASPHOT and GALFIT do not obtain an acceptable fit and the contamination of 
the Bulge dominated objects.
%We point out that the significance of the figure decreases significantly beyond
%$z\sim2$, given the small number of galaxies per bin in that redshift range 
%(see Fig. \ref{dz} for the redshift distribution). 
The discrepancy between GASPHOT and GALFIT for objects with $z\gtrsim2$ 
(classified by the former as late and by the latter as early-type) must be 
considered rather marginal, due to the small number of galaxies in that
redshift range for which GASPHOT and GALFIT found acceptable fits.
 
\subsection{The Evolution of the Merging Fraction}

The high resolution very deep imaging in the K20 field gives us also 
the opportunity to investigate the evolution of the merging fraction
with the redshift for this K-band selected sample. 
Up to date, studies 
of the merging fraction against redshift have been done for 
optically selected (Patton et al. 1997, Le F\`evre et al. 2000)
and NIR selected (Conselice et al. 2003) samples.
In this paragraph, we investigate the evolutionary merging fraction using
both pairs statistics (as i.e. Le F\`evre 2000) and asymmetry technique
(Conselice et al. 2003).

Following Le F\`evre et al. (2000), we have visually identified 
in the K20 sample objects having a companion brighter than i=24.5 within 
20 kpc from the center, independently on the morphology of the main object,
after removing cluster objects.

Given that in almost all cases we have the redshift only for the 
main object and we do not know if the observed pair consists of truly 
interacting objects or not. Then, we need to apply a 
correction factor accounting for this projection contamination. 
This correction is calculated integrating the galaxy number counts published 
by Driver et al. (1995) up to I=24.5 within a circle of physical radius 
of 20 kpc.
Given that at $z<0.5$ the 20 kpc radius projects over a large area, the high 
contamination of spurious pairs makes not statistically significant the
measures at such low redshifts.

A second correction is needed to estimate how many of these 
physical pairs are going to merge. Here we use the correction proposed 
by Patton et al. (1997), suggesting that in the local universe half of 
the pairs with relative velocities less than $350 Km s^{-1}$  are 
expected to merge within 0.5 Gyr.
Given that the velocity is expected to vary with redshift as $(1+z)^{-1}$, 
the merging fraction is obtained by multiplying the pairs fraction 
(corrected for effects of projection) by $0.5(1+z)$.

For comparison, we have recomputed the merging fraction in a completely different way,
based on asymmetry computation (Conselice et al. 2003):
an object is identified as merger if it has an asymmetry greater than a certain
value (in this case $A>0.4$). Also in this case cluster objects have been removed.
This technique selects a different class of objects compared
with those identified by pairs statistics. Given
that it is computed in a small area around the object, the highest asymmetry is
measured for strongly interacting objects, rather than for objects in a pair often
separated by many kiloparsecs, and that therefore do not fall together in the 
area for the asymmetry computation. Moreover, the merging fraction obtained by
the asymmetry technique strongly depends on the border value of A used for
merger identification. 

\begin{figure}
\begin{center}
\includegraphics[width=\columnwidth]{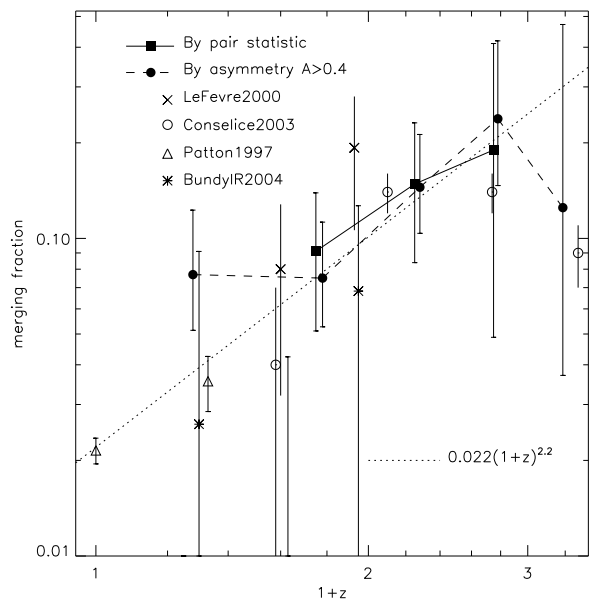}
\end{center}
\caption{
The inferred merging fraction for the K20 sample as a function of the redshift,
both from pairs statistics (filled circles) and from asymmetry criterion (filled
squares).
Errors bar are calculated by a Poisson statistic.
Results of previous works are also reported, together with the best-fit 
relation to the data (dotted line).
}
\label{merging_fraction}
\end{figure}

Figure \ref{merging_fraction} reports the inferred merging fraction for
the K20 sample as a function of the redshift both from pair statistic 
(filled circles) and from asymmetry criterion (filled squares). 
The results from the two techniques are in good agreement with each other,
and with those obtained by other authors. Our data, together
with those by Le F\`evre et al. (2000), Patton et al. (1997) and Conselice et al. 
(2003) provide the evolution of the merging 
fraction as $a(1+z)^m$, where $a$ and $m$ are free parameters:
\begin{equation}
f_{merg}=(0.022\pm0.002)\left(1+z\right)^{2.2\pm0.3}.
\end{equation}
The value of $m$ is in quite good agreement with $m=2.8\pm0.9$ obtained
by Patton et al. (1997) using only its data at $z<0.3$, but is smaller
than the value obtained by Le F\`evre et. al 2000 ($m=3.4\pm0.6$).
We can conclude that the fraction of merging evaluated for a K-band selected
sample increases up to $z\sim1.5$, according with previous works, but,
due to the limited statistic, we can not constrain the evolution for higher
redshift. There are however indications that it remains lower than 20\% up to
z=2.

It must be noted that if it is true that the asymmetry and pairs techniques 
select different merging events (observed respectively in late and early 
phases), the merging fraction could be the sum of the two, then the numbers 
in Fig. \ref{merging_fraction} must be multiplied by a factor 2.

\section{Comparing the morphological and spectroscopic classification}\label{mospe}

In Table \ref{morph_vs_spec_t} we report a comparison between morphological 
and spectroscopic classifications for our sample galaxies. The two 
classifications have been made completely independently.

\begin{table*}%[!ht]
\begin{center}
\begin{tabular}{cccccc|c}
\hline
spectral & & perturbed & normal & perturbed & & all\\
class & ell. & ellipticals &  spirals & spirals & irr/mer & types\\
\hline
Early type     & 51 & 5 & 6  & 3  & 1 & 66\\
Early+EmLines  & 5  & 3 & 20 & 3  & 7 & 38\\
Emission lines & 2  & 6 & 53 & 41 & 76 & 178\\
Unobs./Unid.   & 2  & 0 & 1  & 1  & 14 & 18\\
\vspace{1mm}\\
\hline
Total          & 60 & 14 & 80 & 48 & 98 & 300\\
\hline
\end{tabular}
\end{center}
\caption{
Comparison between morphological and spectroscopical classification.
}\label{morph_vs_spec_t}
\end{table*}

Among the 18 spectroscopically unobserved or unidentified objects, 
15 fall in the irregular class. The bulk of them (9/15) are low 
surface brightness objects very close to the optical detection limits.
For the remaining object with both spectroscopic and morphological
classification, the agreement is quite good: 51/60 normal ellipticals have 
an early type spectrum, 118/128 spirals (morphological class 3 and 4) 
and 97/98 irregulars have late type or early+emission-line spectra. 

Among the galaxies classified as normal ellipticals, there are 5 objects 
with early spectra, but showing some emission lines, and 
2 galaxies with emission lines spectra.
Among these 7 objects, 4 are small and with difficult morphological 
classification (S\'ersic indices $n>2.3$ and $r_e<0.25''$); 
one has been classified as S0/Sa, because of the axial ratio near to 0.5, 
even if a clear disk component is not so evident; 
the remaining 2 have clearly an elliptical morphology ($n>4$). 

Among the 14 galaxies that we have morphologically classified as peculiar
ellipticals (galaxies with a dominant elliptical component, but with signs
for distortion of the isophotes), 9 have emission lines in their spectrum, showing
that a certain degree of activity is really present in the galaxies. 
In this morphological class however 5 galaxies that show no signs 
of emission lines fall.

Ten late-type galaxies (belonging to the morphological classes 3, 
4 and 5) show early-type spectra with no signs for emission lines. 
In 6 cases a luminous and concentrated bulge dominates the spectrum, 
probably overwhelming the contributions from the star-forming regions.  
In one case the classification is uncertain due to the small area covered
by the galaxy.
The agreement between morphological and spectroscopic properties is good also
for the galaxies belonging to the two clusters: only 1 object classified as
elliptical shows signs of emission lines, whereas 4 late-type galaxies have
early-type spectra.

\section{Galaxy sizes versus redshift}\label{sizes}

A critical prediction of galaxy formation models concerns the evolution 
of the physical sizes. The disk scale-lengths, in particular, are predicted 
to decrease proportionally to the inverse of the Hubble parameter (Fall \&
Efstathiou 1980).
We have investigated the evolution with the redshift of the galaxy sizes 
for early/type (class 1 and 2), disks (class 3 and 4) and irregulars 
(class 5) up to $z\sim1.5$.

As a measure of the physical size of galaxies we use the effective 
radius $r_e$ retrieved by GASPHOT by fitting the galaxy light profile 
with a S\'ersic model.

\begin{figure}
\begin{center}
\includegraphics[width=\columnwidth]{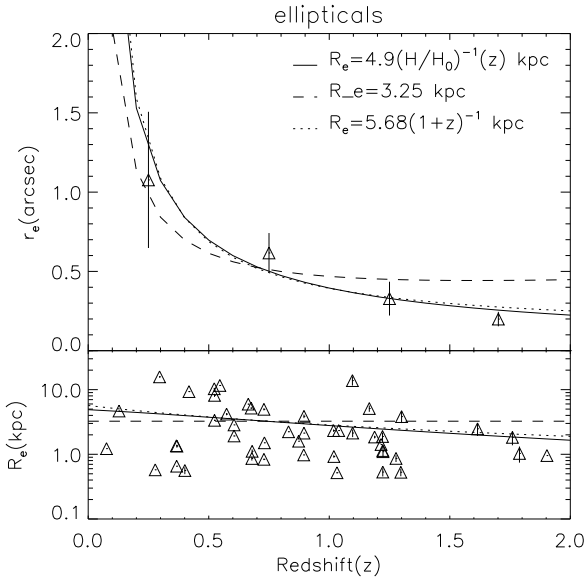}
\end{center}
\caption{
The evolution with the redshift of the galaxy sizes of elliptical class. 
{\it Upper panel}: mean angular size in four redshift bins compared with 
three models of evolution for the physical galaxy size. For each point 
error bars represent the standard deviation divided by the square root 
of the number of galaxies in that bin. Continuous and dotted lines are 
two model in which galaxy size evolves with the redshift respectively 
as $(H/H_0)^{-1}$ and as $(1+z)^{-1}$, whereas dashed line is a model 
of constant size. The three curves are normalized to the observed mean 
value at redshift $0.5<z<1$.
{\it Lower panel}: the physical sizes of early type galaxies are reported 
as a function of the redshift. The three models of evolution and 
no-evolution of galaxy sizes are also reported.
}\label{rzelli}
\end{figure}

\begin{figure}
\begin{center}
\includegraphics[width=\columnwidth]{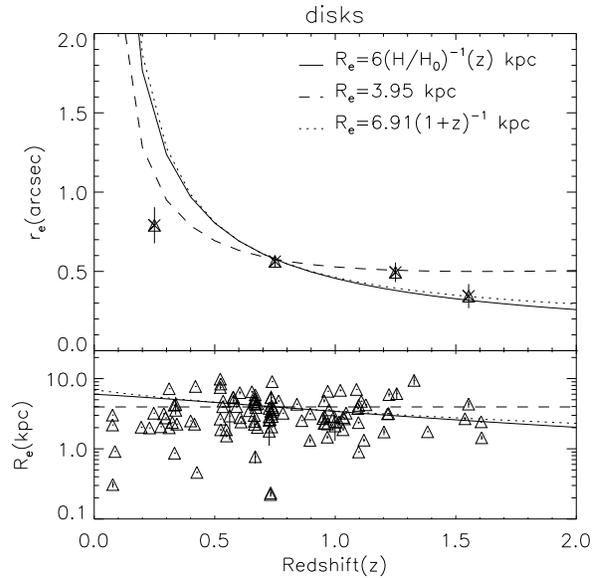}
\end{center}
\caption{
Same as Fig. \ref{rzelli} for disk galaxies.
}\label{rzspir}
\end{figure}

\begin{figure}
\begin{center}
\includegraphics[width=\columnwidth]{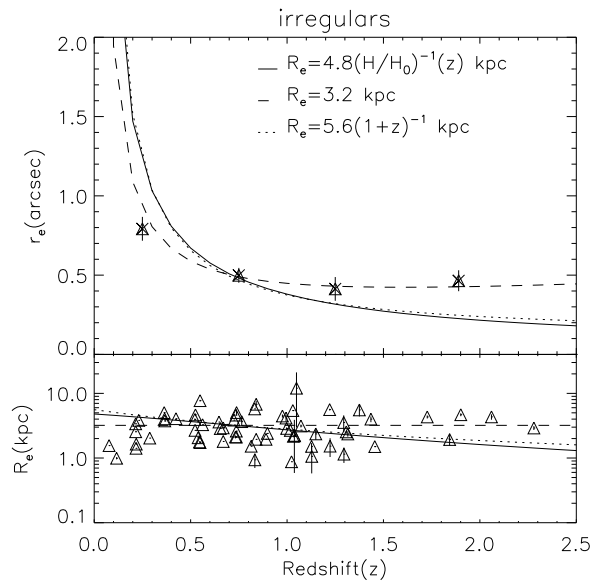}
\end{center}
\caption{
Same as Fig. \ref{rzelli} for irregular galaxies.
}\label{rzirr}
\end{figure}

In the upper panels of Figs. \ref{rzelli}, \ref{rzspir} and \ref{rzirr} 
we report the mean angular size in four redshift bins ($z<0.5$, $0.5<z<1$, 
$1<z<1.5$ and $z>1.5$) respectively for ellipticals, spirals and irregulars. 
Cluster galaxies can be recognized at $\bar{z}=0.736$ and $\bar{z}=0.668$.

For comparison, the curves corresponding to no-evolution of the physical 
size, evolution with the inverse of Hubble parameter $H(t)$ and evolution 
with $(1+z)^{-1}$ are also plotted, each normalized to the observed 
datapoint at $z=0.75$. 
In the lower panels, the data on the physical sizes for individual galaxies 
are reported.

Our results do not show much evidence for size evolution in the explored 
redshift range. In particular, spirals and irregulars are entirely 
consistent with no evolution, in good agreement with the results of 
Ravindranath et al. (2004), who payed particular attention in discussing 
the observational biases.

A small decrease of $R_e$ may be seen for the elliptical population, 
for which our results are consistent with $R_e\propto (H/H_0)^{-1}$. 
The statistical significance of this result is however not at all marginal:
we point out that at high $z$, $K$-band selects only the most massive 
and luminous galaxies, while at low $z$ the explored mass interval is much more wide.

\section{Discussion and conclusions}\label{conclusion}

We have exploited very deep and high resolution ACS/HST 
imaging recently become available in the CDFS area and covering a major portion
of the K20 survey, to conduct an accurate morphological analysis of the K20
sample high-redshift galaxies. For each object, the analysis has been performed in the ACS band closer to
the B-band rest-frame in order to minimize effects of morphological
k-correction.

Our main results are hereafter summarized.

\begin{enumerate}
\item
From visual classification (performed by three authors), we find that: 
60 galaxies are normal ellipticals or S0; 14 are peculiar ellipticals; 
80 are normal spirals; 48 spirals showing signs of interaction or of regions of excess
star formation; 98 are irregular galaxies. 
%Using the multi-band photometric imaging data available, the objects 
%are split into 4 redshift bins, so that each object
%is analyzed roughly in its B-band rest-frame, to minimize effects of 
%the morphological $K$-correction. 
We have carefully investigated the effects
of the morphological $K$-correction on the objects at $z\gtrsim1.2$, for which
the B-band rest-frame is not available, establishing robust upper and lower
limits to the fraction of each morphological class.

\item
We have investigated the capabilities of parametric and non-parametric 
techniques in disentangling morphological classes with a low level 
of interaction.
In particular, we used the GALFIT and GASPHOT packages to fit the light 
distribution with the S\'ersic parametric model, to exploit the correlation
between S\'ersic indices and B/D ratios.

We have shown that parametric tools like GASPHOT and GALFIT are not completely
efficient in distinguishing between early- and late-type objects. They do not
find acceptable fits for $\sim50/300$ objects, mainly those with high asymmetry or
low surface brightness. 
The general agreement of these two tools for the remaining objects is quite good
(while for $\sim10\%$ they turn out to be completely inconsistent). The agreement 
worsens at increased asymmetry of the objects, as expected (see Figs. 
\ref{sersic_gal_vs_gas} and \ref{re_gal_vs_gas}). The distributions
of the S\'ersic indices for the different visual morphological classes turn out
to be quite similar for the two tools (see Fig. \ref{mclas_gal_vs_gas}).
It is interesting to note that a quite substantial number ($\sim50/250$) of visually classified late-type galaxies have a high S\'ersic index ($n>2$): this case may be
produced by a prominent bulge on top of a tiny disk structure.

\item
We have complemented the morphological analysis with the calculation of non-parametric
quantities like Concentration, Asymmetry and Clumpiness for the objects in the 
sample.
We have shown that there is a precise region of the CAS space (defined by Eq. 
\ref{earlycas}) occupied mostly by early type galaxies: ellipticals are the 
objects with higher Concentration and smaller Clumpiness and Asymmetry. 
Only one elliptical falls out, while 20 late type lye within this region. 

Simulations have been performed to investigate the reliability of the CAS
values retrieved at high redshifts. It has been shown that some biases
in the measures are introduced by the degradation of the resolution at
high z, somehow preventing an assessment of the intrinsic evolution
of the CAS parmeters.

\item
We have shown that a criterion based on the S\'ersic index only does not completely
segregate early- by late-type galaxies (see Fig. \ref{frac4_z}; 
we have already mentioned the many low surface brightness or highly asymmetric 
objects for which the fit procedures do not converge, and the number of
late-type galaxies with a S\'ersic index $n_{Sersic}>2$).

The CAS criterion instead better reproduces the classification between early-
and late-type obtained by visual inspection (see Fig. \ref{frac4_z}).

\item
Over the 300 objects of the sample, only 74 turn out to be early type galaxies
(our class 1 and 2). Their redshift distribution is dominated by the cluster at 
$z\sim 0.737$, and the number decreases rapidly for $z>1.5$ following the 
general trend for a rapid decrease at such $z$ for the whole population.

The most numerous galaxies are the irregulars: 1/3 of the objects in the 
sample belongs this class. Their fraction increases 
from $\sim20\%$ at low-$z$ to $\sim60\%$ at $z>1.2$ and is dominant at higher
redshifts. The simultaneous decrease of early and disk galaxies
suggests intuitively that some amounts of high redshift irregular galaxies 
may progressively transform into the local elliptical and spiral galaxy 
population.
A conclusion on this important issue would however require much improved 
statistics in the critical redshift interval of $z\sim1.5$ to 2.
In any case our analysis indicates a predominance of spirals 
and irregulars in K-band selected samples at even moderate depths.

We stress that these results turn out to be rather robust, thanks to 
the use of the B-band rest-frame up to $z\sim 1.2$, (so minimizing effects of
morphological $K$-correction), and to the accurate assessment
of biases introduced by the $K$-correction at higher $z$.

\item
We also analyse the evolution with the redshift of the merging fraction.
We use both the pair statistics technique and the asymmetry criterion $A>0.4$,
in order to measure merging fraction. Although the two techniques likely select 
two different kinds of objects (corresponding to early and late phases of a 
merger), we found that the results of each criterion are in good agreement with 
each other and with previous works.
The inferred merging fraction remains in any case lower than 20\% up to $z=2$.
If we consider that the total merging fraction is the sum of the two, then 
the numbers in Fig. \ref{merging_fraction} are to be multiplied by a factor $\sim 2$.

\item
We have performed a systematic comparison between morphological and spectroscopic classification, showing quite good agreement:  91\% of the sample have
spectral characteristic compatible with morphological ones. There are
some ellipticals (7/60) with a very relaxed morphology showing emission lines 
in their spectra.
The late-type objects showing early-type passive spectra are mainly bulge-dominated spirals,
for which probably the signal from the disk component is overwhelmed by the concentrated 
and luminous bulge.

\item
Finally the redshift dependence of galaxy sizes is investigated for elliptical
(class 1+2), disk (class 3+4) and irregular (class 5) galaxies separately. Our results show no 
evolution for the sizes of disks and irregulars, whereas a small decrease of 
$R_e$ has been found for elliptical galaxies: their sizes seems to vary as
$R_e\propto (H/H_0)^{-1}$.
These results are particulalry important if we consider that we are selecting at the high redshifts the most luminous, massive, hence the largest, of the galaxy population at that cosmic epoch.

\end{enumerate}

\noindent {\bf ACNOWLEDGEMENTS:}
We are grateful to the referee, Roberto Abraham, for careful reading and interesting
comments on the manuscript.

\label{lastpage}

\end{document}